\documentclass[%
 prd,
 twocolumn,
 superscriptaddress,
 numerical,
 showpacs,
 amsmath,amssymb,
 aps,
 nofootinbib,
 longbibliography,
 floatfix
]{revtex4-1}

\usepackage{bm}
\usepackage{bbm}
\usepackage{cancel}
\usepackage{slashed}
\usepackage{graphicx}
\usepackage{combelow} 
\usepackage{mathtools}
\usepackage[usenames,dvipsnames,svgnames,table]{xcolor}
\usepackage[colorlinks=True,linkcolor=blue,citecolor=blue,urlcolor=blue]{hyperref}
\usepackage{braket}

\let\oldciteauthor=\citeauthor
\def\citeauthor#1{\hypersetup{citecolor=black}\oldciteauthor{#1}}

\let\oldciten=\onlinecite
\def\onlinecite#1{\hypersetup{citecolor=blue}\oldciten{#1}}

\let\oldcite=\cite
\def\cite#1{\hypersetup{citecolor=blue}\oldcite{#1}}

\def\bx{{\mathbf{x}}}
\def\bk{{\mathbf{k}}}

\def\bpi{{\boldsymbol{\pi}}}
\def\bg{{\boldsymbol{\gamma}}}
\def\bA{{\boldsymbol{A}}}
\def\bS{{\boldsymbol{S}}}

\newcommand{\beqn}{\begin{eqnarray}}
\newcommand{\eeqn}{\end{eqnarray}}
\newcommand{\beqs}{\begin{subequations}}
\newcommand{\eeqs}{\end{subequations}\\[-2mm]\noindent}
\newcommand{\eq}[1]{(\ref{#1})}

\newcommand{\bs}{\boldsymbol}
\newcommand{\mathfrakE}{{\cal K}}
\definecolor{purple}{rgb}{0.8,0,0.6}
\definecolor{PURPLE}{rgb}{0.8,0,0.6}
\definecolor{orange}{rgb}{1,0.55,0}

\begin{document}

\title{Helical Separation Effect and helical heat transport for Dirac fermions}

\author{Victor E. Ambru\cb{s}}
\affiliation{Department of Physics, West University of Timi\cb{s}oara, Bd.~Vasile P\^arvan 4, Timi\cb{s}oara 300223, Romania}

\author{Maxim N. Chernodub}
\affiliation{Institut Denis Poisson, CNRS UMR 7013, Universit\'e de Tours, 37200 France}
\affiliation{Department of Physics, West University of Timi\cb{s}oara, Bd.~Vasile P\^arvan 4, Timi\cb{s}oara 300223, Romania}

\date{\today}

\begin{abstract}
An ensemble of massless fermions can be characterized by its total helicity charge given by the sum of axial charges of particles minus the sum of axial charges of antiparticles. We show that charged massless fermions develop a dissipationless flow of helicity along the background magnetic field. We dub this transport phenomenon as the Helical Separation Effect (HSE). Contrary to its chiral cousin, the Chiral Separation Effect, the HSE produces the helical current in a neutral plasma in which all chemical potentials vanish. In addition, we uncover the Helical Magnetic Heat Effect which generates a heat flux of Dirac fermions along the magnetic field in the presence of non-vanishing helical charge density. We also discuss possible hydrodynamic modes associated with the HSE in neutral plasma.
\end{abstract}

\maketitle

\section{Introduction}

Massless or nearly massless fermions can be found in various areas of physics. They are present in theories of fundamental interactions and reveal themselves in cosmological models of the early Universe, ultra-hot relativistic plasmas, superfluids, and many other physical systems~\cite{volovik2003universe}. Discoveries of Dirac and Weyl semimetals, where the massless (chiral) fermions manifest as quasiparticle excitations, have opened up intriguing possibilities for experimental verification of relativistic phenomena associated with chiral fermions~\cite{armitage2018weyl}.

The pivotal properties of these excitations are typically linked to their vector (gauge), axial (chiral) and conformal (scale) symmetries. In the case of semimetals, these symmetries impact the electromagnetic~\cite{vazifeh2013electromagnetic,Chernodub:2017jcp}, thermal~\cite{gooth2017experimental,Chernodub:2021nff}, and elastic~\cite{cortijo2015elastic} responses of the materials. The peculiar characteristics of these semimetals are closely tied to the quantum anomaly that disrupts a continuous symmetry of an underlying classical theory due to quantum fluctuations~\cite{Fujikawa:2004cx}. In fundamental field theories, the vivid example is represented by the axial anomaly which gives rise to exotic transport phenomena of quarks, mediated by the evolving gluon fields' topology in the expanding quark-gluon plasma of heavy-ion collisions~\cite{Kharzeev:2013ffa}.

The continuous symmetries of massless fermions, leaving aside their conformal properties, are usually restricted to vector and axial symmetries, which lead to the conservation of vector (electric) and axial (chiral) charges. The excess of the number of particles over antiparticles, regardless of their spins, gives the vector charge. The sign of the spin projection on the momentum of a fermion or an anti-fermion, regardless of their vector charges, determines the chirality of the (anti-)particle in question. However, an ensemble of massless fermions can be characterized by a third, familiar, and, at the same time, intermittently disregarded quantity: helicity. 

The similarity of fermionic helicity to fermion chirality often leads to the identification of these quantities, even though they represent distinct physical properties of fermions~\cite{Pal:2010ih}. The crucial difference -- and the source of the persistent confusion -- is rooted in the fact that the chirality of a single fermion identically coincides with its helicity while the chirality and helicity of a single {\it anti}-fermion are precisely opposite to each other. These properties lead to erroneous generic statements like ``chirality is the same as helicity,'' which is invalid for systems containing both particles and antiparticles. For example, a system made of a fermion and an anti-fermion can reside in a state with double total chirality and vanishing helicity (when the chiralities of the fermion and anti-fermion are the same), or in another state in which the total chirality equals zero, while helicity is non-vanishing (when the chiralities of the fermion and anti-fermion are opposite).

The helicity and chirality numbers coincide only for ensembles of massless fermions that consist only of particles with vector charges of the same sign. However, in various physically relevant environments -- for example, in neutral finite-temperature plasmas or systems at high temperatures -- this condition is not met, and the helicity should be treated on the same footing as chirality. 

The vector, axial, and helical charges form a triad of quantities that characterize an ensemble of massless fermions and anti-fermions. The conservation of the vector $Q_V$, axial $Q_A$ and helical $Q_H$ charges in closed systems of free massless fermions allows us to introduce the thermodynamically conjugated chemical potentials that determine these charges in thermodynamic system. In particular, we introduce the helical chemical potential $\mu_H$ to encode the helicity imbalance, alongside the more familiar vector and axial chemical potentials, $\mu_V$ and $\mu_A$, respectively. As a result, the fermion ensemble is described by the Fermi-Dirac distribution at finite temperature and chemical potentials $\mu_\ell$ associated with the vector ($\ell = V$), axial ($\ell = A$), and helical ($\ell = H$) charges. The presence of the helical charge in interacting fermionic matter affects its thermodynamics, leading to a rich phase diagram~\cite{Chernodub:2020yaf,Wan:2020ffv}, in clear distinction with the thermodynamic effects of the axial charge density~\cite{Chernodub:2011fr,Ruggieri:2011xc}. Moreover, in addition to the standard vector ${\bs j}_{V}$ and axial ${\bs j}_{A}$ currents, the helical degree of freedom allows us to construct the helical current ${\bs j}_H$ which determines how the helicity is transferred in the medium by fermions.

In our paper, we concentrate on anomalous transport phenomena associated with helical properties of massless fermions in the presence of the background magnetic field following our earlier study to the fermion helicity in vortical backgrounds~\cite{Ambrus:2019ayb,Ambrus:2019khr,Ambrus:2020oiw}. The main result of this paper is represented by the following formula for the helical current, which we dub the Helical Separation Effect (HSE):
\begin{align}
  {\bs j}_H = \frac{T \ln 2}{\pi^2} q {\bs B} \qquad \text{for} \quad \mu_V = \mu_A = \mu_H = 0\,,
\label{eq_HSE}
\end{align}
which shows that in the background of a time-independent uniform magnetic field ${\bs B}$, the neutral matter develops a helical current along the direction of the magnetic field. The associated helical magnetic conductivity is proportional to the temperature $T$, implying that the higher the temperature, the stronger the helical current. Equation~\eq{eq_HSE} is invariant under time reversal transformation, $t \to -t$, implying that the generated helical current is a dissipationless quantity.\footnote{Notice that magnetic fields can also generate dissipationless (persistent) electric current in small metallic rings via the Aharonov-Bohm effect~\cite{Byers:1961zz,Deaver:1961zz,Cotescu2016}. This effect, however, has a different nature than the anomalous transport effects of the kind~\eqref{eq_HSE} since the persistent current in a ring (i) emerges in the transverse direction to the magnetic field and (ii) disappears in the thermodynamic limit, when the radius of the ring becomes infinite.}

As the helicity of massless fermions is similar to the zilch quantum number for photons~\cite{Lipkin1964}, the HSE~\eq{eq_HSE} is a counterpart of the Zilch Vortical Effect which shows how the polarization of photons is transferred in the rotating photon gas~\cite{Chernodub:2018era}. We have shown in Ref.~\cite{Ambrus:2020oiw} that the helicity and axial vortical effects play a complementary role that elegantly accounts for the discrepancy between the polarization of particles and antiparticles. Similarly, one can expect that the new HSE will also play a complementary role to the influence of the CSE on the polarization of particles in heavy-ion collisions discussed in Ref.~\cite{Buzzegoli:2022qrr}.

The HSE~\eq{eq_HSE} has a simple generalization to the case of dense matter characterized by the triad of the chemical potentials $\mu_\ell$ ($\ell = V,A,H$):
\begin{align}
  {\bs j}_H = \frac{T \ln 2}{\pi^2} \kappa_\beta({\bs \mu}) q {\bs B}\,,
\label{eq_HSE_2}
\end{align}
where
\begin{align}
 \kappa_\beta({\bs \mu}) = \frac{1}{4\ln 2} \sum_{\substack{a, b, c = \pm 1 \\ abc=1}} \ln \left( 1 + e^{\beta (a\mu_V + b \mu_A + c \mu_H)} \right)\,,
 \label{eq_kappa_beta}
\end{align}
is a thermodynamic factor normalized as $\kappa_\beta({\bs 0}) = 1$.

The name ``Helical Separation Effect''~\eq{eq_HSE} is inspired by its similarity with the Chiral Separation Effect (CSE)~\cite{Son:2004tq}:
\begin{align}
   {\bs j}_A = \frac{\mu_V}{2 \pi^2} q {\bs B}\,.
   \label{eq_CSE}
\end{align}
The CSE implies that the dense matter with $\mu_V \neq 0$ develops the current of axial (chiral) charge $ {\bs j}_A$ along a background magnetic field~${\bs B}$. Contrary to the CSE~\eq{eq_CSE}, the HSE generates the helical current~\eq{eq_HSE} in the absence of matter imbalance (with all chemical potentials vanishing). Notice that, while the strength of the CSE~\eq{eq_CSE} is determined by the axial anomaly~\cite{Son:2004tq}, the relevance of this anomaly to the helical transport~\eq{eq_HSE} is not certain due to the $\ln 2$ factor in \eq{eq_HSE}. A relevant discussion on the existence of possible helical anomalies in the context of the helical vortical conductivity is given in Ref.~\cite{Ambrus:2019ayb,Ambrus:2019khr}. Finally, we mention that the Helical Separation Effect~\eq{eq_HSE} should not be confused with the Helical Magnetic Effects which appear in the context of Weyl semimetals~\cite{kharzeev2018giant} or in hydrodynamic transport~\cite{Yamamoto:2021gts}.

Another important result of our article is that the magnetic field in fermionic matter with nonzero helical density generates a heat current. Namely, the background magnetic field $\bs B$ in the presence of a nonvanishing helical density encoded in the helical chemical potential $\mu_H$, produces a non-zero heat flux $j_\epsilon^i \equiv T^{0i}$ along the axis of the magnetic field:
\begin{equation}
 {\bs j}_\epsilon = \frac{\mu_H T \ln 2}{\pi^2}  q {\bs B}, \qquad 
\label{eq_heat_flux}
\end{equation}
where $T^{0i}$ is an off-diagonal component of the energy momentum tensor for massless Dirac fermions. The Helical Magnetic Heat Effect~\eq{eq_heat_flux} represents the leading term in the strong field limit, $|qB| \gg T^2$, with higher-order corrections in chemical potentials of the order of $O({\bs \mu}^3)$ being neglected. Notice that the heat current~\eq{eq_heat_flux} is unexpectedly proportional to the first power of temperature $T$, contrary to the $T^2$ heat current produced by the mixed axial-gravitational anomaly~\cite{Landsteiner:2011cp}. A similar unusual feature was also noted for the helical vortical effects which produce a heat current in vortical helical matter~\cite{Ambrus:2019khr}.

This paper is organized as follows. In Sec.~\ref{sec_helicity_magnetic}, we introduce the conserved helicity operator for massive Dirac fermions in a magnetic field. A disadvantage of considering massive fermions is that, in addition to the explicit breaking of chirality conservation, they have a frame-dependence in the definition of helicity. Therefore, while discussing the applications of the Helical Separation Effect at the end of the paper, we will turn to the massless fermionic limit, similar to the generic tactics applied to the chiral degrees of freedom. We start our discussion from the discussion of the formal definition of the helicity operator in magnetic field, Sec.~\ref{sec_helicity_magnetic}, which leads us to the construction of the common mode solutions for the helicity operator and the Hamiltonian in Sec.~\ref{sec:modes}. Using the derived eigenmodes, the finite-temperature field theory at finite helical density is developed in Sec.~\ref{sec_finite_temperature} which brings us to the anomalous transport in Sec.~\ref{sec_anomalous_transport}. We derive not only the Helical Separation Effect~\eq{eq_HSE}, but also the corrections of the helical charges to the chiral magnetic and chiral vortical conductivities as well as to the heat transport. The interplay of the fluctuations of the helical charge density and the energy density -- supported by the Helical Separation Effect~\eq{eq_HSE} and the Helical Magnetic Heat Effect~\eq{eq_heat_flux} -- could, in principle, lead to the appearance of new hydrodynamic excitations. In Sec.~\ref{sec_helical_heat} we analyse this possibility for a {\it  neutral} plasma and show that the helical degree of freedom does not generate a new hydrodynamic wave if all chemical potentials vanish. Our conclusions are summarized in the last section. Throughout this paper, Planck units ($\hbar = c = k_B = 1$) and the $(+---)$ metric signature are adopted.

\section{Helicity operator in magnetic field}
\label{sec_helicity_magnetic}

The symmetric Lagrangian for the Dirac field  $\psi$ with mass $M$ coupled to an external electromagnetic potential reads as follows:
\begin{equation}
 \mathcal{L} = \frac{i}{2} \overline{\psi} \overleftrightarrow{\slashed{\partial}} \psi - q \overline{\psi} \slashed{A} \psi - M \overline{\psi} \psi,
 \label{eq_L}
\end{equation}
where we use the standard convention that $q = -e < 0$ is the 
(negative) electron charge. The Dirac equation~\eq{eq_L} can be obtained using the Euler-Lagrange formalism for both $\psi$ and $\overline{\psi}$:
\begin{align}
 (i \slashed{\partial} - q \slashed{A} - M)\psi &= 0,\nonumber\\
 \overline{\psi}(i \overleftarrow{\slashed{\partial}} + q \slashed{A} + M) &= 0.
 \label{eq_dirac}
\end{align}
The equation for $\psi$ can be cast in the form
\begin{equation}
 i \frac{\partial \psi}{\partial t} = H \psi, \qquad 
 H = \gamma^0 (-i \bs{\gamma} \cdot \bs{\nabla} + q \slashed{A} + M),
 \label{eq_H_def}
\end{equation}
where $H$ is the Hamiltonian operator.

For the case of an electron in a magnetic field, the four-potential $A^\mu$ can be taken in the Coulomb gauge,
\begin{equation}
 A^\mu = (0,  \bA),
\end{equation}
where $\bA \equiv \bA(\bx)$ is a time-independent vector field. In this case, the Hamiltonian becomes
\begin{align}
 H &= M \gamma^0 + \gamma^0 \bg\cdot \bpi \nonumber\\
 &= M \gamma^0 + 2\gamma^5 \bS \cdot \bpi,
\end{align}
where 
\begin{equation}
 \bpi = -i \bs{\nabla} - q \bA
\end{equation}
is the generalized momentum of a charged particle in an electromagnetic field and $\bS = \frac{1}{2} \gamma^5 \gamma^0 \bg$ is the spin matrix.
Due to the time-independence of the Hamiltonian, $\partial_t H = 0$, we can take particle normal-mode solutions $U_j$ which are eigenfunctions of the operator $H$:
\begin{equation}
 H U_j = E_j U_j \Rightarrow  U_j = e^{-i E_j t} \widetilde{U}_j,
 \label{eq_eigen_H}
\end{equation}
where $\widetilde{U}_j \equiv \widetilde{U}_j(\bx)$ is a time-independent spinor field and the cumulative symbol $j$ labels distinct solutions to be discussed in detail later. Similarly, the antiparticle solutions of Eq.~\eqref{eq_dirac} can be obtained by charge conjugation, 
\begin{equation}
 V_j = i \gamma^2 
 U_j^*{\Big |}_{q \rightarrow -q} 
 = e^{i E_j t} \widetilde{V}_j,
 \label{eq_V_j}
\end{equation}
with $\widetilde{V}_j \equiv \widetilde{V}_j(\bx) = i \gamma^2 
\widetilde{U}_j^*{\big |}_{q\rightarrow -q}
$. Clearly, $H V_j = -E_j V_j$.

In order to describe the polarization degree of freedom, we employ the helicity operator, defined as the projection of the spin along the generalized momentum:
\begin{equation}
 h = \frac{\bS \cdot \bpi}{\sqrt{H^2 - M^2}}.
 \label{eq_h}
\end{equation}
Taking into account that $[\gamma^5, \bS] = [\gamma^0, \bS] = 0$, as well as $[H, \bpi] = 0$, it is not difficult to show that $[H, \bS \cdot \bpi] = 0$, from which we directly conclude that $H$ commutes with the helicity operator $h$:
\begin{equation}
 [H, h] = 0.\label{eq_h_commutation}
\end{equation}
Writing conversely
\begin{equation}
 h = \frac{\gamma^5}{2} \frac{H - M \gamma^0}{\sqrt{H^2 - M^2}},
\end{equation}
it is easy to see that $h^2 = 1/4$. 

We can thus take the Hamiltonian eigenfunctions $U_j$, $V_j$ to be simultaneously eigenfunctions of $h$, satisfying
\begin{equation}
 h U_j = \lambda_j U_j, \qquad 
 h V_j = \lambda_j V_j,
 \label{eq_eigen_h}
\end{equation}
with the helicity eigenvalues $\lambda_j = \pm 1/2$  such that $\lambda_j^2 = 1/4$. Note that, at the level of the classical theory, the helicity operator $h$ preserves its eigenvalue under charge conjugation, in contradistinction to the Hamiltonian $H$ or chirality $\gamma^5$ (in the case of massless fermions). Using Eq.~\eqref{eq_eigen_H}, the eigenvalue relations~\eqref{eq_eigen_h} become as follows:
\begin{align}
 h_j \widetilde{U}_j &= \lambda_j \widetilde{U}_j, \qquad\ 
 h_j^c \widetilde{V}_j = \lambda_j \widetilde{V}_j, 
\end{align}
where the helicity operator and its conjugated form reduce, respectively, to the following expressions:
\begin{align}
 h_j &= \frac{1}{2\sqrt{E_j^2 - M^2}} \begin{pmatrix}
     0 & E_j + M\\ 
    E_j - M & 0
 \end{pmatrix},\nonumber\\
 h_j^c &= -\frac{1}{2\sqrt{E_j^2 - M^2}} \begin{pmatrix}
     0 & E_j - M\\ 
    E_j + M & 0
 \end{pmatrix}. 
\end{align}
The above equations are automatically solved by taking 
\begin{align}
 \widetilde{U}_j &= \frac{1}{\sqrt{2}}\begin{pmatrix}
     \mathfrakE^+_j \\
     2\lambda_j 
     \mathfrakE^-_j
 \end{pmatrix} \otimes \phi_j, &
 \widetilde{V}_j &= \frac{1}{\sqrt{2}}\begin{pmatrix}
     2\lambda_j \mathfrakE^-_j \\
     -\mathfrakE^+_j
 \end{pmatrix} \otimes \phi_j^c,
\end{align}
where we introduced the notation 
\begin{equation}
 \mathfrakE^\pm_j = \sqrt{1 \pm \frac{M}{E_j}},  
 \label{eq_beautiful_E}
\end{equation}
The Pauli two-spinors $\phi_j$ and $\phi_j^c$ satisfy 
\begin{align}
 \bs{\sigma} \cdot \bpi \phi_j &= 2\lambda_j \sqrt{E_j^2 - M^2} \phi_j, \nonumber\\
 \bs{\sigma} \cdot \bpi \phi^c_j &= 2\lambda_j \sqrt{E_j^2 - M^2} \phi^c_j,
 \label{eq_eigen_h_phij}
\end{align}
being related through charge conjugation:
\begin{equation}
 \phi_j^c = i \sigma^2 \phi_j^* 
 {\Big |}_{q \rightarrow -q}.
\end{equation}

The helicity operator~\eq{eq_h} shares similarity with the zilch quantum number which -- despite a seemingly non-local definition --  is a well-defined quantity associated with the polarization of photons~\cite{Lipkin1964}.

\section{Mode solutions in a constant magnetic field} 
\label{sec:modes}

In this section, we review the eigensystem of a charged spinor particle in the background of the uniform homogeneous magnetic field, with an emphasis on the role of the helical degrees of freedom.

\subsection{Energy levels}

We seek to describe a particle moving in a constant magnetic field, oriented along the $z$ axis:
\begin{equation}
 {\bs B} = B \mathbf{e}_z,
\end{equation}
where $\mathbf{e}_z$ is a unit vector pointing in the positive direction along the $z$ axis and $B$ is a constant real number.
The above magnetic field can be implemented via the asymmetric four-potential
\begin{equation}
 A^\mu = B x \delta^\mu_y, \label{eq_Amu}
\end{equation}
giving rise to the Faraday tensor 
\begin{equation}
 F_{\mu\nu} = \partial_\mu A_\nu - \partial_\nu A_\mu = -B(g_{\mu x} g_{\nu y} - g_{\mu y} g_{\nu x}).\label{eq_Fmunu}
\end{equation}
We note that the symmetric four-potential $A_{\rm sym}^\mu = \frac{1}{2} B (x \delta^\mu_y - y \delta^\mu_x)$ can be obtained from the four-potential $A^\mu$ in Eq.~\eqref{eq_Amu} via the gauge transformation $A^\mu_{\rm sym} = A^\mu + \partial^\mu \Lambda$, with $\Lambda = -\frac{1}{2} xy B$. For simplicity, we will continue the discussion in the context of the asymmetric choice \eqref{eq_Amu}, bearing in mind that the choice of gauge has no impact on the expectation values that form the main objectives of our paper.

It is easy to check that when $A^\mu$ is given by Eq.~\eqref{eq_Amu}, the Hamiltonian commutes with both $P^y = -i \partial_y$ and $P^z = -i \partial_z$, such that $\phi_j$ and $\phi_j^c$ can be taken as eigenfunctions of these operators,
\begin{equation}
 \phi_j = \frac{e^{i p^y_j y + i p^z_j z}}{2\pi}
 \begin{pmatrix}
  f^+_j \\ f^-_j
 \end{pmatrix}, \quad 
 \phi_j^c = \frac{e^{-i p^y_j y - i p^z_j z}}{2\pi}
 \begin{pmatrix}
  f^{c;+}_j \\ f^{c;-}_j
 \end{pmatrix},
 \label{eq_eigen_p}
\end{equation}
where $f_j^\pm \equiv f_j^\pm(x)$ are scalar functions depending only on the transverse spatial coordinate $x$. The charge conjugates $f_j^{c;\pm}$ are related to $f_j^\pm$ via
\begin{equation}
 f^{c;\pm}_j = \pm 
 f^{\mp; *}_j {\Big |}_{q \rightarrow -q}\,.
\end{equation}

Focussing now on the particle modes only, substituting Eq.~\eqref{eq_eigen_p} into Eq.~\eqref{eq_eigen_h_phij} leads to
\begin{multline}
 \begin{pmatrix}
  2\lambda_j \sqrt{E_j^2 - M^2} - p_j^z & i \sqrt{2|qB|} (\partial_{\xi_j} - \frac{\sigma}{2} \xi_j) \\ 
  i \sqrt{2|qB|} (\partial_{\xi_j} + \frac{\sigma}{2} \xi_j) & 2\lambda_j \sqrt{E_j^2 - M^2} + p^z_j
 \end{pmatrix} \begin{pmatrix}
  f^+_j \\ f^-_j
 \end{pmatrix}\\
 = 0,
\end{multline}
where we introduced
\begin{equation}
 \xi_j = \sqrt{2|qB|} \left(x - \frac{\sigma p^y_j}{|qB|} \right), \quad 
 \sigma = {\rm sgn}(qB).
 \label{eq_xisigma_def}
\end{equation}
With the above notation, the functions $f^\pm_j(\xi_j)$ satisfy the following differential equation:
\begin{gather}
 \left(\frac{\partial^2}{\partial \xi_j^2} - \frac{1}{4} \xi_j^2 + \nu_j^\pm + \frac{1}{2} \right) f_j^\pm = 0, \nonumber\\
 \nu_j^\pm = \frac{E_j^2 - M^2 - (p_j^z)^2}{2|qB|} - \frac{1 \mp \sigma}{2}.
\end{gather}
As is well known, the above equation has regular solutions given in terms of the Hermite polynomials when $\nu_j^\pm$ is a non-vanishing integer:
\begin{equation}
 f^\pm_j = \mathcal{C}_j^\pm e^{-\xi_j^2 / 4} H_{\nu_j^\pm} (\xi_j),
\end{equation}
where $\mathcal{C}^\pm_j$ are integration constants. From the above discussion, it is easy to write the quantization relation for the energy:
\begin{equation}
 E_j^2 = M^2 + (p_j^z)^2 + 2 n_j |qB|, \quad 
 \nu^\pm_j = n_j - \frac{1 \mp \sigma}{2}.
 \label{eq_Ej_B}
\end{equation}

Using now the recurrence relations $H'_\nu(z) = \nu H_{\nu -1}(z)$ and $z H_\nu(z) = H_{\nu + 1}(z) + \nu H_{\nu - 1}(z)$, it is possible to derive the relation
\begin{multline}
 \left(\frac{\partial}{\partial \xi_j} - \frac{1 \mp \sigma}{2} \xi_j \right) H_{\nu}(\xi_j) \\
 = \nu \frac{1 \pm \sigma}{2} H_{\nu - 1}(\xi_j) - \frac{1 \mp \sigma}{2} H_{\nu + 1}(\xi_j),
\end{multline}
which implies that
the coefficients $\mathcal{C}^\pm_j$ satisfy the following matrix equation:
\begin{multline}
 \begin{pmatrix}
  2\lambda_j \sqrt{E_j^2 - M^2} - \sigma p_j^z  & -i \sqrt{2|qB|} \\ 
  i n_j \sqrt{2|qB|} & \hspace{-10pt} 2\lambda_j \sqrt{E_j^2 - M^2} + \sigma p_j^z   
 \end{pmatrix}
 \begin{pmatrix}
  \mathcal{C}_j^\sigma \\ \mathcal{C}_j^{-\sigma}
 \end{pmatrix}\\ = 0.
\end{multline}
The above expression gives one relation between the two normalization constants $\mathcal{C}^{\pm}_j$. The second relation allowing these constants to be determined is provided by the requirement of unit norm of the modes $U_j$ under the Dirac inner product,
\begin{align}
 \langle U_j, U_{j'} \rangle &= \int d^3x \overline{U}_j(x) \gamma^0 U_{j'}(x) = \delta(j,j') \nonumber\\
 &\equiv \delta_{n_j n_{j'}} \delta_{\lambda_j \lambda_{j'}} \delta(p_j^y - p_{j'}^y) \delta(p_j^z - p_{j'}^z).
 \label{eq_norm_def}
\end{align}
The integration with respect to $y$ and $z$ leads to the $\delta(p^y_j - p^y_{j'})$ and $\delta(p^z_j - p^z_{j'})$ factors, while the integration with respect to $x$ can be performed using the orthogonality relation for the Hermite polynomials,
\begin{equation}
 \int_{-\infty}^\infty d\xi_j e^{-\xi_j^2 / 2} H_{n_j}(\xi_j) H_{n_j'}(\xi_j) = \sqrt{2\pi} n_j! \delta_{n_j,n_j'}.
 \label{eq_Hermite_ortho}
\end{equation}
In the following subsections, we will discuss independently the cases $n_j = 0$ (the so-called lowest Landau level, LLL) and $n_j  > 0$ (the higher Landau levels, HLL).

\subsection{Lowest Landau level (LLL)}

On the lowest Landau level (LLL) corresponding to $n_j = 0$, we avoid the divergent solution with $\nu_j^{-\sigma} = -1$ by setting 
\begin{equation}
 \mathcal{C}_j^{-\sigma} {\Big |}_{n_j = 0} = 0.
\end{equation}
The only non-trivial solution corresponds to the polarization satisfying 
\begin{equation}
 2\sigma \lambda_j = {\rm sgn}(p_j^z),
\end{equation}
where we remind the reader that $\sigma = {\rm sgn}(q B)$, cf.~Eq.~\eqref{eq_xisigma_def}.

Normalizing now the modes using the Dirac inner product in Eq.~\eqref{eq_norm_def}, we find the normalization $\mathcal{C}_j^\sigma = (|qB| /\pi)^{1/4}$ corresponding to the LLL mode:
\begin{align}
 U_j(x){\Bigl |}_{n_j = 0} &= \theta(\sigma \lambda_j p_j^z) \frac{e^{-i E_j t}}{\sqrt{2}} 
 \begin{pmatrix}
  \mathfrakE^+_j \\ 2\lambda_j \mathfrakE^-_j
 \end{pmatrix} \otimes \phi^\sigma_j(\bx) {\Bigl |}_{n_j = 0},\nonumber\\
 \phi^\sigma_j{\Bigl |}_{n_j = 0}(\bx) &= 
 \frac{e^{i p_j^y y + i p_j^z z}}{2\pi}
 e^{-\frac{\xi_j^2}{4}} \left(\frac{|qB|}{\pi}\right)^{1/4} 
 \begin{pmatrix}
    \frac{1 + \sigma}{2} \\[1mm]
    \frac{1 - \sigma}{2}
 \end{pmatrix}.
 \label{eq_LLL_U}
\end{align}
Similarly, the antiparticle modes can be found as 
\begin{align}
 V_j(x){\Bigl |}_{n_j = 0} &= \theta(-\sigma \lambda_j p_j^z) \frac{e^{i E_j t}}{\sqrt{2}} 
 \begin{pmatrix}
  2\lambda_j \mathfrakE^-_j \\ -\mathfrakE^+_j
 \end{pmatrix} \otimes \phi^{c;\sigma}_j(\bx){\Bigl |}_{n_j = 0},\nonumber\\
 \phi^{c;\sigma}_j(\bx){\Bigl |}_{n_j = 0} &= 
 \frac{e^{-i p_j^y y - i p_j^z z}}{2\pi}
 e^{-\frac{\xi_{j;c}^2}{4}} \left(\frac{|qB|}{\pi}\right)^{1/4} 
 \begin{pmatrix}
    \phantom{-} \frac{1 + \sigma}{2} \\[1mm]
    -\frac{1 - \sigma}{2}
 \end{pmatrix},\label{eq_LLL_V}
\end{align}
where $\xi_{j;c}$ is the charge conjugate of $\xi_j$,
\begin{equation}
 \xi_{j;c} = 
 \xi_j{\Bigl |}_{q \rightarrow -q} = 
 \sqrt{2|qB|} \left(x + \frac{\sigma p^y_j}{|qB|}\right).
\end{equation}
These expressions allow us to establish a link between $\phi^{c;\sigma}_j$ and its charge conjugate as follows:
\begin{equation}
 \phi^{c;\sigma}_{0,p^y, p^z, \lambda}(\bx) = \sigma \phi^\sigma_{0,-p^y,-p^z, \lambda}(\bx).
\label{eq_relation_phi0}
\end{equation}

\subsection{Higher energy levels}

For $n_j > 0$, we have 
\begin{equation}
 \mathcal{C}^{-\sigma}_j = -\frac{i n_j \sqrt{2|qB|}}{2\lambda_j \sqrt{E_j^2 - M^2} + \sigma p^z_j} \mathcal{C}^\sigma_j,
\end{equation}
which leads to 
\begin{align}
 \mathcal{C}^\sigma_j &= \frac{\mathcal{N}_j}{\sqrt{n_j!}} \left(\frac{|qB|}{4\pi}\right)^{1/4} \mathfrak{p}^{2\sigma \lambda_j}_j, \nonumber\\
 \mathcal{C}^{-\sigma}_j &= \frac{-i \mathcal{N}_j \sqrt{n_j}}{2\lambda_j \sqrt{n_j!}} \left(\frac{|qB|}{4\pi}\right)^{1/4} \mathfrak{p}^{-2\sigma \lambda_j}_j,
\end{align}
where $\mathcal{N}_j$ is a phase factor, $|\mathcal{N}_j| = 1$, while
\begin{equation}
 \mathfrak{p}^{\pm}_j = \sqrt{1 \pm \frac{p_j^z}{\sqrt{E_j^2 - M^2}}}.
\end{equation}
Specifically, we have 
\begin{subequations}\label{eq_n>0_U}
\begin{equation}
 U_j(x) = \frac{e^{-i E_j t}}{\sqrt{2}} \begin{pmatrix}
  \mathfrakE^+_j \\ 2\lambda_j \mathfrakE^-_j
 \end{pmatrix} \otimes \phi^\sigma_j(\bx),
\end{equation}
where the Pauli spinors are given by
\begin{align} 
 \phi^+_j(\bx) &= 
 \frac{e^{i p_j^y y + i p_j^z z}}{2\pi \sqrt{n_j!}}
 e^{-\xi_j^2/4} \left(\frac{|qB|}{4\pi}\right)^{1/4}\nonumber\\
 & \times 
 \begin{pmatrix}
    \mathfrak{p}^{2\lambda_j}_j H_{n_j}(\xi_j) \\ 
    -2i \lambda_j \mathfrak{p}^{-2\lambda_j}_j \sqrt{n_j}  H_{n_j-1}(\xi_j)
 \end{pmatrix},\nonumber\\
 \phi^-_j(\bx) &= 
 \frac{e^{i p_j^y y + i p_j^z z}}{2\pi \sqrt{n_j!}}
 e^{-\xi_j^2/4} \left(\frac{|qB|}{4\pi}\right)^{1/4} \nonumber\\
 &\times 
 \begin{pmatrix}
  \mathfrak{p}^{2\lambda_j}_j \sqrt{n_j} H_{n_j-1}(\xi_j) \\ 
    2i \lambda_j \mathfrak{p}^{-2\lambda_j}_j H_{n_j}(\xi_j)
 \end{pmatrix}.
\end{align}
\end{subequations}
In the above, we used $\mathcal{N}_j = 1$ for $\sigma = 1$ and $\mathcal{N}_j = 2i \lambda_j$ for $\sigma = -1$.
The antiparticle modes 
$V_j = i \gamma^2 U_j^*{\bigl |}_{q \rightarrow -q}$ 
can be obtained through the charge conjugation operation, being given explicitly by
\begin{subequations}\label{eq_n>0_V}
\begin{align}
 V_j(x) &= \frac{e^{i E_j t}}{\sqrt{2}} \begin{pmatrix}
  2\lambda_j \mathfrakE^-_j \\ -\mathfrakE^+_j
 \end{pmatrix} \otimes \phi^{c;\sigma}_j(\bx), \nonumber\\
 \phi^{c;+}_j(\bx) &= 
 \frac{e^{-i p_j^y y - i p_j^z z}}{2\pi \sqrt{n_j!}}
 e^{-\xi_{j;c}^2/4} \left(\frac{|qB|}{4\pi}\right)^{1/4}\nonumber\\
 & \times 
 \begin{pmatrix}
    -2i\lambda_j \mathfrak{p}^{-2\lambda_j}_j H_{n_j}(\xi_{j;c}) \\ 
    -\mathfrak{p}^{2\lambda_j}_j \sqrt{n_j}  H_{n_j-1}(\xi_{j;c})
 \end{pmatrix},\nonumber\\
 \phi^{c;-}_j(\bx) &= 
 \frac{e^{-i p_j^y y - i p_j^z z}}{2\pi \sqrt{n_j!}}
 e^{-\xi^2_{j;c}/4} \left(\frac{|qB|}{4\pi}\right)^{1/4} \nonumber\\
 &\times 
 \begin{pmatrix}
  2i\lambda_j \mathfrak{p}^{-2\lambda_j}_j \sqrt{n_j} H_{n_j-1}(\xi_{j;c}) \\ 
  - \mathfrak{p}^{2\lambda_j}_j H_{n_j}(\xi_{j;c})
 \end{pmatrix}.
\end{align}
It can be seen that the charge-conjugated Pauli spinors $\phi^{c;\sigma}_j(\bx)$ corresponding to higher levels satisfy the relation
\begin{equation}
 \phi^{c;\sigma}_{n,p^y,p^z,\lambda}(\bx) = -2i\lambda \sigma \phi^\sigma_{n,-p^y,-p^z,\lambda}(\bx),
\end{equation}
\end{subequations}
which shares certain similarity with Eq.~\eq{eq_relation_phi0}.

\section{Finite-temperature field theory at finite helical density}
\label{sec_finite_temperature}

\subsection{Second quantization}

The full solution of the Dirac equation reads 
\begin{equation}
 \psi(x) = \sum_j [U_j(x) a_j + V_j(x) b^\dagger_j],
 \label{eq_psi}
\end{equation}
where the symbol $j$, which labels the eigenmodes, collectively denotes the eigenvalues $p^y_j$, $p^z_j$, $n_j$ and $\lambda_j$. The sum over the eigenmodes is shorthanded as follows: 
\begin{equation}
    \sum_j = \sum_{\lambda_j = \pm \frac{1}{2}} \int_{-\infty}^\infty dp^y_j \int_{-\infty}^\infty dp^z_j \sum_{n_j = 0}^\infty,
\end{equation}
where the $n_j = 0$ term corresponds to the contribution coming from the lowest Landau level (LLL), which exists only when $2\sigma \lambda_j p^z_j > 0$. 

Demanding that the equality \eqref{eq_psi} is exact and noting that $a_j= \braket{ U_j, \psi }$ and $b_j^\dagger = \braket{V_j, \psi }$, we arrive to the completeness relation for the set of modes $U_j$ and $V_j$:
\begin{equation}
 \sum_j [U_j(t, \bx) U^\dagger_j(t, \bx') + V_j(t, \bx) V_j^\dagger(t, \bx')] = \delta^3(\bx - \bx').
\end{equation}

Promoting now $\psi(x)$ to a Fock-space operator $\hat{\psi}$, the canonical anti-commutation relation
\begin{equation}
 \{\hat{\psi}(t, \bx), \hat{\psi}^\dagger(t, \bx')\} = \delta^3(\bx - \bx'),
\end{equation}
can be achieved by imposing
\begin{equation}
 \{\hat{a}^\dagger_j, \hat{a}_{j'} \} = \{\hat{b}^\dagger_j, \hat{b}_{j'} \} = \delta(j,j'),
\end{equation}
where $\delta(j,j') = \delta(p^y_j - p^y_{j'}) \delta(p^z_j - p^z_{j'}) \delta_{n_j,n_{j'}} \delta_{\lambda_j, \lambda_{j'}}$, as in Eq.~\eqref{eq_norm_def}.

\subsection{Conserved charges}

In this paper, we describe fermionic ensembles at finite temperature in the background of a constant homogeneous magnetic field. We will investigate states which exhibit imbalance with respect to the electric and helical charges, as well as with respect to the axial charge (in the case of massless fermions). These conserved charges can be obtained by considering the vector, axial and helical charge currents, defined, respectively, as follows:
\begin{align}
 J^\mu_V & = \overline{\psi} \gamma^\mu \psi\,, \nonumber \\ 
 J^\mu_A & = \overline{\psi} \gamma^\mu \gamma^5 \psi\,, \\
 J^\mu_H & = \overline{\psi} \gamma^\mu h \psi + \overline{h \psi} \gamma^\mu \psi\,.  \nonumber 
\end{align}
We now compute the divergences of the above currents. In the case of the vector current, we have
\begin{equation}
 \partial_\mu J^\mu_V = \overline{\slashed{\partial} \psi} \psi + \overline{\psi} \slashed{\partial} \psi.
 \label{eq_div_JV}
\end{equation}
Using the Dirac equation~\eqref{eq_dirac}, the derivatives $\slashed{\partial}\psi$ and $\overline{\slashed{\partial} \psi}$ can be replaced via 
\begin{align}
 \slashed{\partial}\psi &= -i (q \slashed{A} + M) \psi, &
 \overline{\slashed{\partial} \psi} &= i \overline{\psi} (q \slashed{A} + M).
\end{align}
Substituting the above into Eq.~\eqref{eq_div_JV} gives the conservation of the vector current, $\partial_\mu J^\mu_V = 0$. 

Employing the same steps as above for the derivative of the axial four-current $\partial_\mu J^\mu_A$ reveals the partial conservation of the axial current (PCAC),
\begin{equation}
 \partial_\mu J^\mu_A = 2M (i \overline{\psi} \gamma^5 \psi),
 \label{eq_PCAC}
\end{equation}
where the expression between the parentheses on the right-hand side is the pseudoscalar condensate. 

Finally, in the case of the helicity current, we have 
\begin{equation}
 \partial_\mu J^\mu_H = \overline{\slashed{\partial} \psi} h \psi + \overline{\psi} \slashed{\partial} h \psi + \text{h.c.},
\end{equation}
where h.c. denotes the Hermitian conjugate. Taking advantage of the commutation relation \eqref{eq_h_commutation}, it is easy to show that if $\psi$ is a solution of the Dirac equation, then $h \psi$ also satisfies the Dirac equation. This leads to the relation
\begin{equation}
 \overline{\psi} \slashed{\partial} h \psi = -i \overline{\psi} (q \slashed{A} + M) h \psi
 = -\overline{\slashed{\partial} \psi} h \psi,
\end{equation}
implying that the helical current is identically conserved even for massive fermions:
\begin{align}
\partial_\mu J^\mu_H = 0.
\label{eq_d_JH}
\end{align}
Note however that the conservation of the helicity current is broken perturbatively by helicity-violating pair annihilation processes, as discussed in Ref.~\cite{Ambrus:2019khr}. In a setting of real physical systems, such as the quark-gluon plasma, the non-conservation of helicity due to perturbative scattering has the same timescale as the non-conservation of chirality due to dynamical mass generation {\cite{Astrakhantsev:2019zkr}, which allows us to treat the helicity and chirality on equal basis~\cite{Ambrus:2019khr}.

The above discussions prompt us to introduce the total vector, axial and helical charges,
\begin{align}
 Q_{\ell} = \int d^3x\, J^0_{\ell}\,, \qquad \ell = V,A,H\,,
\end{align}
which are identically conserved for the vector and helical degrees of freedom:
\begin{align}
  \partial_t Q_V = \partial_t Q_H = 0\qquad\text{[for fermions with any mass]},
\end{align}
for both chiral and massive fermions. However, due to the PCAC property~\eq{eq_PCAC}, the axial charge does not represent a conserved quantity for a massive particle:
\begin{equation}
 \partial_t Q_A = 2M \eta, \quad 
 \eta \equiv \int d^3x\, i\overline{\psi} \gamma^5 \psi.
 \label{eq_d_QA}
\end{equation}
Thus, while at finite mass $M \neq 0$, the axial charge $Q_A$ is not a good conserved number, the vector and helical charges are identically conserved.

Thanks to the helicity eigenmode solutions discussed in Sec.~\ref{sec:modes}, both the vector and the helicity charge operators $\widehat{Q}_{V/H}$ are diagonal and, together with the Hamiltonian $\widehat{H}$, they admit the following Fock space representation:
\begin{align}
 :\widehat{H}: &= \sum_j E_j (\hat{a}^\dagger_j \hat{a}_j + \hat{b}^\dagger_j \hat{b}_j), \nonumber\\
 :\widehat{Q}_V: &= \sum_j (\hat{a}^\dagger_j \hat{a}_j - \hat{b}^\dagger_j \hat{b}_j), \nonumber\\
 :\widehat{Q}_H: &= \sum_j 2\lambda_j (\hat{a}^\dagger_j \hat{a}_j - \hat{b}^\dagger_j \hat{b}_j),
 \label{eq_HQ_decomposition}
\end{align}
where the colons 
$:\widehat{{\mathcal O}}: \equiv \widehat{{\mathcal O}} - \braket{0|\widehat{{\mathcal O}}|0}$
denote Wick (normal) ordering of an operator $\widehat{\mathcal O}$. 
It is worth pointing out that, while $\widehat{H}$ is even under charge conjugation (both particles and antiparticles contribute with the same sign to $\widehat{H}$), both the vector and the helical charge operators are odd and therefore discriminate between particles and antiparticles. The CPT properties of $\widehat{Q}_H$ are discussed in Ref.~\cite{Ambrus:2019ayb}.

As a side remark, the Hamiltonian $H$ can be obtained from the Dirac energy-momentum tensor, $\Theta^{\mu\nu}$, defined via Noether's theorem as
\begin{equation}
 \Theta^{\mu\nu} = \frac{\partial \mathcal{L}}{\partial(\partial_\mu \psi)} \partial^\nu \psi + \partial^\nu \overline{\psi} \frac{\partial \mathcal{L}}{\partial(\partial_\mu \overline{\psi})} - g^{\mu\nu} \mathcal{L},
 \label{eq_Theta_general}
\end{equation}
where the Dirac Lagrangian $\mathcal{L}$ is given in Eq.~\eqref{eq_L}. Taking into account that $\mathcal{L}$ vanishes when $\psi$ satisfies the Dirac equation, we arrive at the following relation for the energy-momentum tensor:
\begin{equation}
 \Theta^{\mu\nu} = \frac{i}{2} \overline{\psi} \gamma^\mu \overleftrightarrow{\partial^\nu} \psi.
 \label{eq_Theta}
\end{equation}
Due to the presence of interactions, $\Theta^{\mu\nu}$ is, in general, not conserved:
\begin{align}
\partial_\mu \Theta^{\mu\nu} = q J^\mu_V \partial^\nu A_\mu.
\label{eq_d_energy}
\end{align}
The right-hand side of the above expression is not gauge invariant. To restore gauge invariance, it is customary to alter $\Theta^{\mu\nu}$ by adding a gauge-fixing total divergence term, $\Delta \Theta^{\mu\nu} = \partial_\rho (F^{\mu\rho} A^\nu) = -qJ_V^\mu A^\nu + F^{\mu\rho} \partial_\rho A^\nu$, where we used Maxwell's equations, $\partial_\mu F^{\mu\nu} = q J_V^\nu$. Adding the first term, $-q J^\mu A^\nu$, to $\Theta^{\mu\nu}$ yields $\widetilde{\Theta}^{\mu\nu} = \frac{i}{2} \gamma^\mu \overleftrightarrow{\partial^\nu} \psi - qJ^\mu A^\nu$. Its divergence, $\partial_\mu \widetilde{\Theta}^{\mu\nu} = -q J_{V;\mu} F^{\mu\nu}$, represents the relativistic Lorentz force acting on the electric charge and is a gauge-invariant quantity. The second term, $F^{\mu\rho} \partial_\rho A^\nu$, serves to make the electromagnetic contribution to the energy-momentum tensor, $\Theta^{\mu\nu}_{\rm e.m.} = -F^{\mu\lambda} \partial^\nu A_\lambda$, as well as its divergence, gauge-invariant. However, the gauge-fixing term becomes position-dependent, since $A^\nu = B x \delta^\nu_y$, as shown in Eq.~\eqref{eq_Amu}. For this reason, we will not consider the gauge-fixing term and we will focus instead on the form of the energy-momentum tensor $\Theta^{\mu\nu}$ given in Eq.~\eqref{eq_Theta}.
 
The static homogeneous magnetic field considered in this paper, $A_\mu = B x g_{\mu y}$, does not produce work as the Lorentz force is always normal to the direction of the current. Therefore, for the energy component of the conservation relation~\eq{eq_d_energy}, $\nu = 0$, we have $\partial_\mu \Theta^{\mu 0} = 0,$ so that the energy is conserved, as expected. This property allows the Hamiltonian to be defined as the total conserved energy, 
\begin{equation}
 H = \int d^3x\, \Theta^{00}.
\end{equation}
Upgrading $H$ to the Fock space operator $\widehat{H}$ and employing the explicit mode decomposition \eqref{eq_psi} gives the expression in Eq.~\eqref{eq_HQ_decomposition}.

Finally, in the case of chiral (massless) fermions, the axial charge is conserved, $\partial_t Q_A = 0$, and the corresponding Fock-space operator $\widehat{Q}_A$ admits the decomposition
\begin{equation}
 :\widehat{Q}_A: = \sum_j 2\lambda_j (\hat{a}^\dagger_j \hat{a}_j + \hat{b}^\dagger_j \hat{b}_j)
 \qquad \text{[for  $M=0$]}.
\end{equation}
Note that, contrary to the vector and helical charge operators $\widehat{Q}_V$ and $\widehat{Q}_H$, the axial charge operator $\widehat{Q}_A$ is even with respect to charge conjugation (see Table~1 in Ref.~\cite{Ambrus:2019khr}).

\subsection{Thermal expectation values}

We now construct thermal states at finite vector and helical chemical potentials, using the statistical operator~\cite{Canuto:1968apg,Canuto:1968nzn,Canuto:1968tav}
\begin{equation}
 \hat{\rho} = e^{-\beta(:\widehat{H}: - :\boldsymbol{\mu} \cdot \widehat{\mathbf{Q}}:)},
 \label{eq_rho}
\end{equation}
where $\widehat{H}$ is the Hamiltonian, while the conserved charges sector contains the vector and helical charges for massive fermions. In the case of massless fermions, we allow also for a chiral charge, so that we introduce a generic shift of the Fermi energy levels controlled by the whole triad of the corresponding chemical potentials:
\begin{equation}
 \boldsymbol{\mu} \cdot \widehat{\mathbf{Q}} = 
 \mu_V \widehat{Q}_V  + \mu_A \widehat{Q}_A + \mu_H \widehat{Q}_H\,.
\label{eq_bold_mu}
\end{equation}
One should bear in mind that the axial chemical potential $\mu_A$ vanishes for a massive particle, since the axial charge is not conserved if the particle mass is nonzero, cf.~\eq{eq_d_QA}.

The statistical operator \eqref{eq_rho} describes a system at rest, having four-velocity 
\begin{equation}
 u^\mu \partial_\mu = \partial_t.\label{eq_umu}
\end{equation}
With the above four-velocity vector, we can construct the covariant magnetic field four-vector,
\begin{equation}
 B^\mu = \frac{1}{2} \varepsilon^{\mu\nu\alpha\beta} u_\nu F_{\alpha\beta}= B \delta^\mu_z,
\end{equation}
where $\varepsilon^{\mu\nu\alpha\beta}$ is the Levi-Civitta tensor and we took the sign convention such that $\varepsilon^{0123} = 1$.

The thermal expectation value of an operator $\widehat{A}$ is defined as
\begin{equation}
 \braket{\widehat{A}} \equiv \mathcal{Z}^{-1} {\rm tr}(\hat{\rho} \widehat{A}), \qquad 
 \mathcal{Z} = {\rm tr}(\hat{\rho}).
\end{equation}
In the following, we focus only on operators which are quadratic with respect to the field operator $\hat{\psi}$, thus allowing for the following decomposition:
\begin{multline}
 \widehat{A} = \sum_{j,j'} \left[ \mathcal{A}(U_j, U_{j'}) \hat{a}^\dagger_j \hat{a}_{j'} + \mathcal{A}(V_j, V_{j'}) \hat{b}_j \hat{b}^\dagger_{j'} \right.\\
 \left. + \mathcal{A}(U_j, V_{j'}) \hat{a}^\dagger_j \hat{b}^\dagger_{j'} + \mathcal{A}(V_j, U_{j'}) \hat{b}_j \hat{a}_{j'} \right],
 \label{eq_A_generic}
\end{multline}
where $\mathcal{A}(\psi,\chi)$ is a sesquilinear form in the sense that 
\begin{align}
 \mathcal{A}(\alpha \psi + \beta \phi,\chi) &= \alpha^* \mathcal{A}(\psi, \chi) + \beta^* \mathcal{A}(\phi, \chi), \nonumber\\
 \mathcal{A}(\psi,\alpha \chi + \beta \phi) &= \alpha \mathcal{A}(\psi, \chi) + \beta \mathcal{A}(\psi, \phi).
\end{align}
Evaluating $\braket{\widehat{A}}$ essentially boils down to evaluating the thermal expectation values of quadratic products of the one-particle operators, which we discuss below.

First, we note that the decompositions \eqref{eq_HQ_decomposition} imply the following commutation relations:
\begin{align}
 [\widehat{H}, \hat{a}^\dagger_j] &= E_j \hat{a}^\dagger_j, & 
 [\widehat{H}, \hat{b}^\dagger_j] &= E_j \hat{b}^\dagger_J,\nonumber\\
 [\widehat{Q}_V, \hat{a}^\dagger_j] &= \hat{a}^\dagger_j, &  
 [\widehat{Q}_V, \hat{b}^\dagger_j] &= -\hat{b}^\dagger_j, \nonumber\\
 [\widehat{Q}_H, \hat{a}^\dagger_j] &= 2\lambda_j \hat{a}^\dagger_j, & 
 [\widehat{Q}_H, \hat{b}^\dagger_j] &= -2\lambda_j \hat{b}^\dagger_j.
\end{align}
A fourth line can be added for massless fermions, namely 
\begin{equation}
 M = 0: \quad 
 [\widehat{Q}_A, \hat{a}^\dagger_j] = 2\lambda_j \hat{a}^\dagger_j, \quad 
 [\widehat{Q}_A, \hat{b}^\dagger_j] = 2\lambda_j \hat{b}^\dagger_j. 
\end{equation}
With the above relations, it can be established that
\begin{equation}
 \hat{\rho} \hat{a}^\dagger_j \hat{\rho}^{-1} = 
 e^{-\beta \mathcal{E}_j^+} \hat{a}^\dagger_j,\quad 
 \hat{\rho} \hat{b}^\dagger_j \hat{\rho}^{-1} = 
 e^{-\beta \mathcal{E}_j^-} \hat{b}^\dagger_j,
\end{equation}
where we introduced the shifted energy levels
\begin{equation}
 \mathcal{E}_j^{\varsigma_j} = E_j - \varsigma_j \mu_V - 2 \lambda_j \mu_A - 2 \varsigma_j \lambda_j \mu_H,
 \label{eq_Ecal}
\end{equation}
where the quantum number  $\varsigma_j = \pm 1$ distinguishes particles and antiparticles:
\begin{equation}
\varsigma_j = 
\left\{
\begin{tabular}{rcl}
$+1$, & \quad\ & \text{for particles},\\[1mm]
$-1$, & \quad\ & \text{for antiparticles}.
\end{tabular}
\right.
\end{equation}
Equation~\eq{eq_Ecal} takes into account the difference in the positions of the Fermi levels for particles with vector, axial, and helical charges which are anchored in the corresponding chemical potentials $\mu_V$, $\mu_A$, and $\mu_H$, respectively. Despite the seemingly complicated form of this equation, it has a simple meaning, consistent with Eq.~\eq{eq_bold_mu}: the vector chemical potential $\mu_V$ controls the excess of particles over antiparticles which, in turn, are distinguished by the quantum number $\varsigma_j$. The axial chemical potential $\mu_A$ applies equally to particles and antiparticles but distinguishes the polarization states described by $\lambda_j$. Finally, the helical chemical potential $\mu_H$ is sensitive to the helicity of the particle which is opposite for particles and antiparticles, hence the product $\varsigma_j \lambda_j$ in the last term of Eq.~\eq{eq_Ecal}.

The thermal expectation value of the quadratic form $\hat{a}^\dagger_j \hat{a}_{j'}$ thus evaluates to
\begin{align}
 \langle \hat{a}^\dagger_j \hat{a}_{j'} \rangle &= 
 e^{-\beta \mathcal{E}^+_j} \langle \hat{a}_{j'} \hat{a}^\dagger_j \rangle
 = \frac{\delta(j,j')}{e^{\beta \mathcal{E}^+_j} + 1},
\end{align}
where on the last line we used $\braket{\hat{a}_{j'} \hat{a}^\dagger_j} = \delta(j',j) - \braket{\hat{a}^\dagger_j \hat{a}_{j'}}$. Similarly,
\begin{align}
 \langle \hat{b}^\dagger_j \hat{b}_{j'} \rangle &= 
 \frac{\delta(j,j')}{e^{\beta \mathcal{E}^-_j} + 1}.
\end{align}
Finally, it is not difficult to see that 
\begin{equation}
 \braket{\hat{a}_j^\dagger \hat{b}^\dagger_{j'}} = \braket{\hat{b}_j \hat{a}_{j'}} = 0.
\end{equation}

We are now in a position to evaluate the thermal expectation value of the operator $\widehat{A}$ in Eq.~\eqref{eq_A_generic}. Focussing on thermal effects, we consider the expectation value of the normal-ordered operator $:\widehat{A}: = \widehat{A} - \braket{0|\widehat{A}|0}$, which evaluates to
\begin{equation}
 A \equiv \braket{:\widehat{A}:} = \sum_j 
 \left[\frac{\mathcal{A}(U_j, U_j)}{e^{\beta \mathcal{E}^+_j} + 1} - 
 \frac{\mathcal{A}(V_j, V_j)}{e^{\beta \mathcal{E}^-_j} + 1}\right].
 \label{eq_A_tev}
\end{equation}
For simplicity, we will often consider separately the contributions due to the LLL ($n_j = 0$) and to the higher-energy modes ($n_j > 0$) by writing 
\begin{equation}
 A = A_0 + \sum_{n_j = 1}^\infty A_{n_j},
\end{equation}
where $A_0$ corresponds to the lowest Landau level (with $n_j = 0$). The terms $A_{n_j}$ are obtained after performing the $p^y_j$ and $p^z_j$ integrations, as well as the sum over polarization $\lambda_j$:
\begin{equation}
 A_{n_j} = \sum_{\lambda_j} \int dp^y_j \int dp^z_j \left[\frac{\mathcal{A}(U_j, U_j)}{e^{\beta \mathcal{E}^+_j} + 1} - 
 \frac{\mathcal{A}(V_j, V_j)}{e^{\beta \mathcal{E}^-_j} + 1}\right].
\end{equation}

\subsection{Scalar and pseudoscalar condensates} 

Let us apply the previously-discussed formalism to the computation of the scalar and pseudoscalar condensates, $\bar{\psi} \psi$ and $i \bar{\psi} \gamma^5 \psi$. In the latter case, it is not difficult to see that the mode solutions in Eqs.~\eqref{eq_LLL_U} and \eqref{eq_n>0_U} give vanishing contributions to the pseudoscalar condensate,
\begin{equation}
 i \overline{U}_j \gamma^5 U_j = 0,
\end{equation}
and similarly $i \overline{V}_j \gamma^5 V_j = 0$.

In the case of the scalar condensate, the antiparticle contribution can be obtained via charge conjugation, as follows:
\begin{equation}
 \overline{V}_j V_j = -(\overline{U}_j U_j)^*_{q \rightarrow -q}.
\end{equation}
On the LLL, we have 
\begin{equation}
 \overline{U}_j U_j{\Big|}_{n_j = 0} = \frac{M e^{-\xi_j^2 / 2}}{4\pi^2 E_j} \sqrt{\frac{|qB|}{\pi}} \theta(\sigma \lambda_j p^z_j),
\end{equation}
while for $n_j > 0$, 
\begin{multline}
 \overline{U}_j U_j {\Big|}_{n_j > 0} = \frac{M e^{-\xi_j^2 / 2}}{8 \pi^2 E_j n_j!} \sqrt{\frac{|qB|}{\pi}} \\\times 
 (\mathfrak{p}_{j;2\sigma\lambda_j}^2 H_{n_j}^2 + n_j \mathfrak{p}_{j;-2\sigma\lambda_j}^2 H^2_{n_j - 1}),
\end{multline}
where it is understood that the Hermite polynomials take the argument $\xi_j$. In order to evaluate the chiral condensate $\braket{\bar{\psi}\psi}$, the integration with respect to $p^y_j$ can be performed using the orthogonality relation \eqref{eq_Hermite_ortho} for the Hermite polynomials,
\begin{equation}
 \int_{-\infty}^\infty dp^y\, e^{-\xi^2/2} H_n(\xi) H_m(\xi) = \sqrt{\pi |qB|}\, n! \delta_{nm}.
 \label{eq_intpy}
\end{equation}
Specifically, we have 
\begin{equation}
 \overline{U}_j U_j = \frac{M|qB|}{4\pi^2 E_j} 
 \begin{cases}
     \theta(\sigma\lambda_j p^z_j), & n_j = 0,\\
     1, & n_j > 0.
 \end{cases}
\end{equation}
The final result reads
\begin{equation}
 \bar{\psi} \psi = \frac{M|qB|}{4\pi^2} \sum_{n_j = 0}^\infty \sum_{\lambda_j, \varsigma_j} \int_0^\infty \frac{dp_j^z\, g_j}{E_j(e^{\beta\mathcal{E}_j} + 1)},
\end{equation}
where the summation runs over the helicity quantum number $\lambda_j = \pm 1/2$, particle-antiparticle number $\varsigma_j = \pm 1$ and $\mathcal{E}_j \equiv \mathcal{E}^{\varsigma_j}_j$ was introduced in Eq.~\eqref{eq_Ecal}. The degeneracy factor evaluates to $g_j = 1$ for $n_j = 0$ and $g_j = 2$ for $n_j > 0$.

\subsection{Charge currents}

The sesquilinear forms corresponding to the charge current operators read
\begin{align}
 \mathcal{J}^\mu_V(\psi,\chi) & = \overline{\psi} \gamma^\mu \chi, \nonumber\\
 \mathcal{J}^\mu_A(\psi,\chi) & = \overline{\psi} \gamma^\mu \gamma^5 \chi, \\
 \mathcal{J}^\mu_H(\psi,\chi) & = \overline{\psi} \gamma^\mu h \chi + \overline{h \psi} \gamma^\mu \chi\,. \nonumber
\end{align}
Replacing now $\psi = \chi = U_j$ or $\psi = \chi = V_j$ in terms of the helicity eigenmodes discussed in Sec.~\ref{sec:modes}, it can be seen that the sesquilinear forms for the helical charge current can be obtained from the ones for the vector charge current via
\begin{align}
 \mathcal{J}^\mu_H(U_j, U_j) &= 2\lambda_j \mathcal{J}^\mu_V(U_j, U_j), \nonumber\\
 \mathcal{J}^\mu_H(V_j, V_j) &= 2\lambda_j \mathcal{J}^\mu_V(V_j, V_j).
\end{align}
The expectation values $J^\mu_\ell(x) \equiv \braket{:\widehat{J}^\mu_\ell(x):}$ with $\ell \in \{V, A, H\}$ can be evaluated using Eq.~\eqref{eq_A_tev}, where the sesquilinear forms $\mathcal{J}^\mu_\ell(V_j, V_j)$ corresponding to the antiparticle modes can be obtained from those corresponding to the particle modes by using the following properties:
\begin{align}
 \overline{V}_j \gamma^\mu V_j &= (\overline{U}_j \gamma^\mu U_j)^*_{q\rightarrow -q}, \nonumber\\
 \overline{V}_j \gamma^\mu \gamma^5 V_j &= -(\overline{U}_j \gamma^\mu \gamma^5 U_j)^*_{q\rightarrow -q}.
\end{align}
In the following, we will focus on the explicit computation of the sesquilinear forms corresponding to the particle modes, which we will denote in shorthand notation as
\begin{equation}
 \mathcal{J}^\mu_{\ell ;j} \equiv \mathcal{J}^\mu_\ell (U_j, U_j).
\end{equation}

On the LLL, an explicit computation shows that 
\begin{equation}
 \left.\phi^{\sigma;\dagger}_j \sigma^x \phi^\sigma_j\right|_{n_j = 0} = \left.\phi^{\sigma;\dagger}_j \sigma^y \phi^\sigma_j\right|_{n_j = 0} = 0,
\end{equation}
such that 
\begin{equation}
 \left.\mathcal{J}^x_{\ell ;j}\right|_{n_j = 0} = \left.\mathcal{J}^y_{\ell ;j}\right|_{n_j = 0} = 0.
\end{equation}
The temporal and vertical components can be computed as follows:
\begin{align}
 \left.\mathcal{J}^t_{V;j} \right|_{n_j = 0} &= \frac{e^{-\xi_j^2 / 2}}{4\pi^2} \sqrt{\frac{|qB|}{\pi}} \theta(\sigma \lambda_j p_j^z),\nonumber\\ 
 \left.\mathcal{J}^z_{V;j}\right|_{n_j = 0} &= \frac{2\lambda_j \sigma |p_j^z| e^{-\xi_j^2 / 2}}{4\pi^2 E_j} \sqrt{\frac{|qB|}{\pi}} \theta(\sigma \lambda_j p_j^z), \nonumber\\
 \left. \mathcal{J}^t_{A;j}\right|_{n_j=0} &= \frac{2\lambda_j |p_j^z| e^{-\xi_j^2 / 2}}{4\pi^2 E_j} \sqrt{\frac{|qB|}{\pi}} \theta(\sigma \lambda_j p_j^z),\nonumber\\ 
 \left. \mathcal{J}^z_{A;j}\right|_{n_j=0} &= \frac{\sigma e^{-\xi_j^2 / 2}}{4\pi^2}  \sqrt{\frac{|qB|}{\pi}} \theta(\sigma \lambda_j p_j^z). 
 \label{eq_Jmu_sesqui}
\end{align}
Using Eq.~\eqref{eq_intpy}, the $p^y_j$ integral can be evaluated, leading to
\begin{align}
 \begin{pmatrix}
  J^t_{V;0} \\
  J^t_{A;0} \\
  J^t_{H;0}
 \end{pmatrix} &= \frac{|qB|}{4\pi^2} \sum_{\lambda_j, \varsigma_j}
 \int_0^\infty \frac{dp_j^z}{e^{\beta \mathcal{E}_j} + 1}
 \begin{pmatrix} 
  \varsigma_j \\ 
  2\lambda_j p_j^z / E_j \\ 
  2\lambda_j \varsigma_j  
 \end{pmatrix}, \nonumber\\
 \begin{pmatrix}
  J^z_{V;0} \\
  J^z_{A;0} \\
  J^z_{H;0}
 \end{pmatrix} &= \frac{qB}{4\pi^2} \sum_{\lambda_j,\varsigma_j}
 \int_0^\infty \frac{dp_j^z}{e^{\beta \mathcal{E}_j} + 1}
 \begin{pmatrix} 
  2\lambda_j p_j^z / E_j \\  
  \varsigma_j \\
  p^z_j / E_j
 \end{pmatrix},
\end{align}
where $E_j = \sqrt{M^2 + p_{z;j}^2}$ on the LLL.


At higher Landau levels with $n_j > 0$, we compute $\mathcal{J}^\mu_{\ell ;j}$
as follows:
\begin{align}
 \mathcal{J}_{V;j}^t &= \frac{e^{-\xi_j^2/2}}{8\pi^2 n_j!} \sqrt{\frac{|qB|}{\pi}} (\mathfrak{p}^{j;2}_{2\sigma\lambda_j} H_{n_j}^2 
 + \mathfrak{p}^{j;2}_{-2\sigma\lambda_j} n_j H_{n_j-1}^2), \nonumber\\
 \mathcal{J}_{V;j}^y &= -\frac{qB e^{-\xi_j^2/2}}{4\pi^2 E_j(n_j-1)!} \sqrt{\frac{2}{\pi}} H_{n_j} H_{n_j - 1}, \nonumber\\
 \mathcal{J}_{A;j}^z &= \frac{\sigma e^{-\xi_j^2/2}}{8\pi^2 n_j!} \sqrt{\frac{|qB|}{\pi}} (\mathfrak{p}^{j;2}_{2\sigma\lambda_j} H_{n_j}^2 - \mathfrak{p}^{j;2}_{-2\sigma\lambda_j} n_j H_{n_j-1}^2),\nonumber\\
 \begin{pmatrix}
  \mathcal{J}^t_{A;j} \\ \mathcal{J}^y_{A;j} \\ \mathcal{J}^z_{V;j} 
 \end{pmatrix} &= 2\lambda_j \sqrt{1 - \frac{M^2}{E_j^2}} 
 \begin{pmatrix}
  \mathcal{J}^t_{V;j} \\ \mathcal{J}^y_{V;j} \\ \mathcal{J}^z_{A;j} 
 \end{pmatrix},
\end{align}
while $\mathcal{J}_{V;j}^x = \mathcal{J}_{A;j}^x = 0$. As before, the Hermite polynomials take the argument $\xi_j$.
Performing the $p^y_j$ integral using Eq.~\eqref{eq_intpy} leads to
\begin{align}
 \int dp^y_j 
 \begin{pmatrix}
  \mathcal{J}_{V;j}^t \\
  \mathcal{J}_{A;j}^t \\
  \mathcal{J}_{H;j}^t
 \end{pmatrix}
 &= \frac{|qB|}{4\pi^2}
 \begin{pmatrix} 
  1 \\
  2\lambda_j \mathfrakE_j^+ \mathfrakE_j^- \\
  2\lambda_j
 \end{pmatrix}, \nonumber\\
 \int dp^y_j \mathcal{J}_{\ell ;j}^x &= 
 \int dp^y_j \mathcal{J}_{\ell ;j}^y = 0,\nonumber\\
 \int dp^y_j  \begin{pmatrix}
  \mathcal{J}_{V;j}^z\\
  \mathcal{J}_{A;j}^z \\
  \mathcal{J}_{H;j}^z
 \end{pmatrix}
 &= \frac{p^z_j |qB|}{4\pi^2 E_j} 
 \begin{pmatrix}
  1 \\ 
  2\lambda_j / \mathfrakE^+_j \mathfrakE^-_j \\
  2\lambda_j
 \end{pmatrix},
\end{align}
where the quantities $\mathfrakE^\pm_j$ are given in Eq.~\eq{eq_beautiful_E}. It can be seen that $\mathcal{J}^z_{\ell ;j}$  are odd with respect to $p^z_j \rightarrow -p^z_j$ when $n_j > 0$, thus making vanishing contributions to the corresponding terms $J^z_{\ell ;n_j}$:
\begin{align}
 \begin{pmatrix} 
  J^t_{V;n_j>0} \\ J^t_{A;n_j>0} \\ J^t_{H;n_j>0}
 \end{pmatrix} &= \frac{|qB|}{2\pi^2} \sum_{\lambda_j, \varsigma_j}
 \int_0^\infty \frac{dp_j^z}{e^{\beta \mathcal{E}_j} + 1}
 \begin{pmatrix} 
  \varsigma_j \\ 
  2\lambda_j \mathfrakE^+_j \mathfrakE^-_j \\ 
  2\lambda_j \varsigma_j  
 \end{pmatrix}, \nonumber\\
 J^z_{\ell ;n_j>0} &= 0.
\end{align}

\subsection{Energy-momentum tensor}

In the case of the energy-momentum tensor $\Theta^{\mu\nu}$, the sesquilinear forms $\mathcal{T}^{\mu\nu}(V_j,V_j)$ corresponding to the antiparticle modes can be related to those pertinent to the particle modes via
\begin{equation}
 \mathcal{T}^{\mu\nu}(V_j,V_j) = -[\mathcal{T}^{\mu\nu}(U_j,U_j)]^*_{q \rightarrow -q}.
\end{equation}
The sesquilinear forms $\mathcal{T}^{\mu\nu}_j \equiv \mathcal{T}^{\mu\nu}(U_j, U_j)$ corresponding to the non-vanishing entries of the canonical energy-momentum tensor \eqref{eq_Theta} can be written in terms of those encountered for the vector current,
\begin{gather}
 \mathcal{T}^{tt}_j = E_j \mathcal{J}^t_{V;j}, \quad 
 \mathcal{T}^{yy}_j = p^y_j \mathcal{J}^y_{V;j}, \quad 
 \mathcal{T}^{zz}_j = p^z_j \mathcal{J}^{z}_{V;j}, \nonumber\\
 \mathcal{T}^{tz}_j = p^z_j \mathcal{J}^t_{V;j}, \qquad 
 \mathcal{T}^{zt}_j = E_j \mathcal{J}^z_{V;j}.
\end{gather}
The other components vanish, except for
$\mathcal{T}^{xx}_j$, which receives non-vanishing contributions only for $n_j > 0$:
\begin{align}
 \mathcal{T}^{xx}_j &= -\frac{i}{2} \sqrt{2|qB|}\, \overline{U}_j \gamma^x \overleftrightarrow{\partial}_{\!\!\xi_j} U_j \nonumber\\
 &= \frac{|qB|^{3/2} e^{-\xi_j^2/2}}{4\pi^2 E_j (n_j - 1)! \sqrt{\pi}}W_{\xi_j}(H_{n_j-1}, H_{n_j}),
\end{align}
where $W_z(f,g) \equiv f(z) g'(z) - f'(z) g(z)$ is the Wronskian. Due to symmetry constraints, we expect that $\Theta^{xx} = \Theta^{yy}$. 

On the LLL, when $n_j = 0$, we have $\mathcal{T}^{xx}_j \rfloor_{n_j = 0} = \mathcal{T}^{yy}_j \rfloor_{n_j = 0} = 0$, while the non-vanishing components of $\Theta^{\mu\nu}_0$ can be obtained as
\begin{align}
 \Theta^{tt}_0 &= \frac{|qB|}{4\pi^2} \sum_{\lambda_j, \varsigma_j} \int_0^\infty \frac{dp_j^z E_j}{e^{\beta \mathcal{E}_j} + 1}, \nonumber\\
 \Theta^{tz}_0 = \Theta^{zt}_0 &= \frac{qB}{4\pi^2} \sum_{\lambda_j,\varsigma_j} 2 \varsigma_j\lambda_j \int_0^\infty \frac{dp_j^z p_j^z}{e^{\beta\mathcal{E}_j} + 1},\nonumber\\
 \Theta^{zz}_0 &= \frac{|qB|}{4\pi^2} \sum_{\lambda_j,\varsigma_j} \int_0^\infty \frac{dp_j^z p_{z;j}^2}{E_j(e^{\beta\mathcal{E}_j} + 1)}.
\end{align}

In order to compute the contributions due to the higher energy levels  ($n_j > 0$), we replace $p^y_j = x q B - \sigma \xi_j \sqrt{|qB| / 2}$ and use the relation $H_{n_j}'(\xi_j) = n_j H_{n_j-1}(\xi_j)$, such that 
\begin{align}
 \int_{-\infty}^\infty dp^y_j \mathcal{T}^{tt}_j{\Bigl|}_{n_j > 0} &= \frac{E_j |qB|}{4\pi^2}, \nonumber\\
 \int_{-\infty}^\infty dp^y_j \mathcal{T}^{tz}_j{\Bigl|}_{n_j > 0} &=  \int_{-\infty}^\infty dp^y_j \mathcal{T}^{zt}_j{\Bigl|}_{n_j > 0} =
 \frac{p^z_j |qB|}{4\pi^2}, \nonumber\\
 \int_{-\infty}^\infty dp^y_j \mathcal{T}^{xx}_j{\Bigl|}_{n_j > 0} &= \int_{-\infty}^\infty dp^y_j \mathcal{T}^{yy}_j{\Bigl|}_{n_j > 0} = \frac{n_j (qB)^2}{4\pi^2 E_j},\nonumber\\
 \int_{-\infty}^\infty dp^y_j \mathcal{T}^{zz}_j{\Bigl|}_{n_j > 0} &= \frac{(p^z_j)^2 |qB|}{4\pi^2 E_j}.
\end{align}
It can be seen that $\mathcal{T}^{tz}_j|_{n_j>0} = \mathcal{T}^{zt}_j|_{n_j>0}$ are odd with respect to $p_j^z \rightarrow -p_j^z$, thus $\Theta^{tz}_{n_j} = \Theta^{zt}_{n_j} = 0$ for $n_j > 0$. 
The nonvanishing components of the contribution of the higher Landau levels to the expectation value of the energy-momentum tensor $\Theta^{\mu\nu}_{n_j > 0}$ are given by
\begin{align}
 \Theta^{tt}_{n_j>0} &= \frac{|qB|}{2\pi^2} 
 \sum_{\lambda_j,\varsigma_j} \int_0^\infty \frac{dp_j^z E_j}{e^{\beta \mathcal{E}_j} + 1}, \nonumber\\
 \Theta^{xx}_{n_j>0} = \Theta^{yy}_{n_j>0} &= 
 \frac{n_j |qB|^2}{2\pi^2} \sum_{\lambda_j,\varsigma_j} 
 \int_0^\infty \frac{dp_j^z}{E_j(e^{\beta \mathcal{E}_j} + 1)},\nonumber\\
 \Theta^{zz}_{n_j>0} &= \frac{|qB|}{2\pi^2} \sum_{\lambda_j,\varsigma_j} \int_0^\infty \frac{dp_j^z\, p_{z;j}^2}{E_j(e^{\beta \mathcal{E}_j} + 1)}.
\end{align}

\section{Anomalous transport}
\label{sec_anomalous_transport}

For a classical (non-quantum) fluid in thermodynamic equilibrium, the charge currents $J^\mu_\ell$ and energy-momentum tensor $\Theta^{\mu\nu}$ take the ideal fluid form,
\begin{equation}
 J^\mu_{\ell ;{\rm ideal}} = Q_\ell u^\mu, \quad \Theta^{\mu\nu}_{\rm ideal} = \epsilon u^\mu u^\nu - P \Delta^{\mu\nu},
\end{equation}
where $u^\mu$ is the local fluid four-velocity, given by Eq.~\eqref{eq_umu} for the present case, while $\Delta^{\mu\nu} = g^{\mu\nu} - u^\mu u^\nu$ is the projector on the hypersurface orthogonal to $u^\mu$. The thermodynamic pressure $P \equiv P(T, \mu_\ell) = -\frac{1}{3} \Delta_{\mu\nu} \Theta_{\rm ideal}^{\mu\nu}$ is related to the charge densities $Q_\ell = u_\mu J^\mu_\ell$ and the entropy density $s$ via
\begin{equation}
 Q_\ell = \frac{\partial P}{\partial \mu_\ell}, \qquad\
 s = \frac{\partial P}{\partial T},
\end{equation}
while the energy density $\epsilon = u_\mu T^{\mu\nu} u_\nu$ is given via the Euler relation by
\begin{equation}
 \epsilon = sT - P + \bs{\mu} \cdot\mathbf{Q}.
\end{equation}

The purpose of this section is to establish the expressions for the quantities appearing above in the presence of a constant magnetic field, as well as to highlight deviations from the perfect fluid form, which we will interpret as a signature of an anomalous transport phenomenon associated with helical, rather than axial degrees of freedom. It is worth mentioning that the term ``anomalous transport'' denotes dissipationless -- often off-equilibrium -- transport effects that appear due to anomalous breaking of a continuous symmetry. The CSE~\eq{eq_CSE} results from the anomalous breaking of the axial symmetry~\cite{Son:2004tq}. The presence of an anomaly for helical degrees of freedom has been speculated in Refs.~\cite{Ambrus:2019ayb,Ambrus:2019khr} and below we will show that the magnetic field produces the Helical Separation Effect which appears to be very similar to its famous chiral magnetic counterpart, the CME.

\subsection{Hydrodynamic decomposition}
\label{sec:anomalous_transport:hydro}

In general, the charge currents $J^\mu_\ell$ ($\ell \in \{V, A, H\}$) can be decomposed with respect to the fluid four-velocity $u^\mu$ as follows:
\begin{equation}
 J_\ell^\mu = Q_\ell u^\mu + j_\ell^\mu, \qquad 
 Q_\ell = J^t_\ell,
\end{equation}
where $j^\mu_\ell = \Delta^\mu_\alpha J^\alpha_\ell$ represents the charge flow in the fluid rest frame. Since the only vector orthogonal to $u^\mu$ available in this problem is $B^\mu$, we have the following charge flows:
\begin{equation}
 j^\mu_\ell = \sigma^B_\ell B^\mu, \label{eq_j_B}
\end{equation}
with 
\begin{equation}
\sigma^B_\ell = \frac{1}{B} J^z_\ell,
\end{equation}
being the magnetic conductivity of the $\ell$'th charge.

We now consider the fluid described by a symmetric energy-momentum tensor $\Theta^{\mu\nu}$.
Once the velocity $u^\mu$ is fixed (we take $u^\mu = \delta^\mu_t$), $\Theta^{\mu\nu}$ can be decomposed as 
\begin{equation}
 \Theta^{\mu\nu} = \epsilon u^\mu u^\nu - (P + \Pi) \Delta^{\mu\nu} + \pi^{\mu\nu} + j_\epsilon^\mu u^\nu + u^\mu j_\epsilon^\nu.
 \label{eq:Delta4}
\end{equation}
Besides the energy density $\epsilon$ and thermodynamic pressure $P$, the above decomposition introduces as deviations from the ideal fluid form the dynamic (bulk) pressure $\Pi =-P -\frac{1}{3} \Delta_{\mu\nu} \Theta^{\mu\nu}$, the shear-stress tensor $\pi^{\mu\nu} = \Delta^{\mu\nu}_{\alpha\beta} \Theta^{\alpha\beta}$ and the heat flux in the fluid rest frame $j_\epsilon^\mu \equiv \Delta^\mu_\alpha u_\beta \Theta^{\alpha\beta}$. In the above, we employed the projector $\Delta^{\mu\nu}_{\alpha\beta}$ defined as
\begin{equation}
 \Delta^{\mu\nu}_{\alpha\beta} = \frac{1}{2}(\Delta^\mu_\alpha \Delta^\nu_\beta + \Delta^\mu_\beta \Delta^\nu_\alpha) - \frac{1}{3} \Delta^{\mu\nu} \Delta_{\alpha\beta}.
\end{equation}
Specializing the above decomposition to the present case with $u^\mu \partial_\mu = \partial_t$, the scalar quantities $\epsilon$, $P$ and $\Pi$ can be obtained as
\begin{equation}
 \epsilon = T^{tt}, \qquad P + \Pi = \frac{1}{3} (\Theta^{xx} + \Theta^{yy} + \Theta^{zz}).
 \label{eq:hydro_eps_P_plus_Pi}
\end{equation}
In classical fluids, the dynamic pressure $\Pi$ is usually related to the fluid expansion rate $\theta \equiv \partial_\mu u^\mu$, which vanishes in the present case. However, as we will see later, the Dirac fluid in a magnetic field exhibits non-vanishing dynamic pressure. 

As in the case of the charge flow $j^\mu_\ell$, the requirement of orthogonality to the fluid velocity $u^\mu$ restricts the heat flux $j_\epsilon^\mu$ to the form
\begin{equation}
 j_\epsilon^\mu = \sigma_\epsilon^B B^\mu \qquad  \Rightarrow \qquad
 \sigma_\epsilon^B = \frac{1}{B} \Theta^{tz},
\end{equation}
which we dub the Helical Magnetic Heat Effect [cf. Eq.~\eq{eq_heat_flux}].
The shear-stress tensor $\pi^{\mu\nu}$ can be written in terms of the available tensors of the problem,
\begin{equation}
 \pi^{\mu\nu} = a u^\mu u^\nu + b B^\mu B^\nu + c (u^\mu B^\nu + u^\nu B^\mu) + d \Delta^{\mu\nu},
\end{equation}
where $a$, $b$, $c$, and $d$ are certain coefficients.
Imposing $\pi^{\mu\nu} u_\nu = a u^\mu + c B^\mu = 0$ reveals that $a = c = 0$. Furthermore, the tracelessness condition $\pi^\mu{}_\mu = a - b B^2 + 3d = 0$ gives $b = 3d / B^2$, such that 
\begin{align}
 \pi^{\mu\nu} &= d \left(\Delta^{\mu\nu} + 3 \frac{B^\mu B^\nu}{B^2}\right) \nonumber\\
 &= \pi_B \times {\rm diag}\left(0,-\frac{1}{2}, -\frac{1}{2}, 1\right),
 \label{eq:hydro_pimunu}
\end{align}
where we defined $\pi_B = 2d$ for notational convenience.

Before ending this section, we remind the reader that in non-equilibrium relativistic hydrodynamics, the mass-energy equivalence gives rise to an ambiguity in defining the energy density, in the sense that the heat flux can be in part or totally absorbed into a new timelike vector that plays the role of the fluid velocity. In the above discussion, we have implicitly employed the so-called beta (or thermometer) velocity frame \cite{van12,van13,landsteiner13lnp,becattini15epjc}, which gives the velocity in terms of the temperature four-vector defining the density operator $\hat{\rho}$. Another frequently used hydrodynamic frame is the so-called Landau (or energy) frame, defined by the eigenvalue equation 
\begin{equation}
    \Theta^\mu{}_\nu  u^\nu_{\rm L} = \epsilon_{\rm L} u^\mu_{\rm L}.
\end{equation}
In the present case, it is not difficult to establish that
\begin{equation}
 u^\mu_{\rm L} \partial_\mu = \Gamma_{\rm L} (\partial_t + \beta_{\rm L} \partial_z), \qquad 
 \beta_{\rm L} = \frac{\Theta^{tz}}{\epsilon_{\rm L} + \Theta^{zz}},
\end{equation}
where $\Gamma_{\rm L} = (1 - \beta_{\rm L})^{-1/2}$ is the Lorentz factor of the Landau velocity and the Landau energy density is 
\begin{equation}
 \epsilon_{\rm L} = \frac{1}{2} \left[\Theta^{tt} - \Theta^{zz} + \sqrt{(\Theta^{tt} + \Theta^{zz})^2 - 4\Theta_{tz}^2}\right].
\end{equation}
It is easy to see that if $\Theta^{tz} \neq 0$ then, in general, $\epsilon_{\rm L} \neq \epsilon$.

\subsection{Thermometer frame analysis}
\label{sec:anomalous_transport:beta_frame}

Let us now write down the full expressions for the charge densities $Q_\ell = J^t_\ell$, energy density $\epsilon = \Theta^{tt}$ and isotropic pressure computed in the thermometer frame:
\begin{align}
 \begin{pmatrix} 
  Q_V \\ Q_A \\ Q_H 
 \end{pmatrix} &= \frac{|qB|}{4\pi^2} \sum_{n_j = 0}^\infty \sum_{\lambda_j, \varsigma_j}
 \int_0^\infty \frac{dp_j^z\, g_j}{e^{\beta \mathcal{E}_j} + 1}
 \begin{pmatrix} 
  \varsigma_j \\ 
  2\lambda_j \mathfrakE_j^+ \mathfrakE_j^- \\ 
  2\lambda_j \varsigma_j 
 \end{pmatrix},\nonumber\\
 \epsilon &= \frac{|qB|}{4\pi^2} \sum_{n_j = 0}^\infty \sum_{\lambda_j, \varsigma_j}
 \int_0^\infty \frac{dp_j^z\, g_j E_j}{e^{\beta \mathcal{E}_j} + 1},\nonumber\\
 P + \Pi &= \frac{|qB|}{12\pi^2} \sum_{n_j = 0}^\infty \sum_{\lambda_j, \varsigma_j}
 \int_0^\infty \frac{dp_j^z\, g_j (E_j^2 - M^2)}{E_j(e^{\beta \mathcal{E}_j} + 1)},
 \label{eq:thermo_macro}
\end{align}
where $g_j = 1$ for $n_j = 0$ and $g_j = 2$ for $n_j > 0$, $\mathfrakE_j^\pm = \sqrt{1 \pm M^2 / E_j^2}$ [cf. Eq.~\eq{eq_beautiful_E}], and $E_j \sqrt{p_{z;j}^2 + M^2 + 2n_j |qB|}$ as in Eq.~\eq{eq_Ej_B}, while $\mathcal{E}_j = E_j - \varsigma_j \mu_V - 2\lambda_j \mu_A - 2\lambda_j \varsigma_j \mu_H$ according to Eq.~\eq{eq_Ecal}.
Similarly, $\pi_B = \Theta^{zz} - (P + \Pi)$ can be obtained as
\begin{equation}
 \pi_B = \frac{|qB|}{6\pi^2} \sum_{n_j = 0}^\infty \sum_{\lambda_j, \varsigma_j}
 \int_0^\infty \frac{dp_j^z\, g_j (p_{z;j}^2 - n_j |qB|)}{E_j(e^{\beta \mathcal{E}_j} + 1)}.\label{eq:thermo_piB}
\end{equation}

The above expressions do not explicitly distinguish between the LLL and the higher energy-level contributions. On the contrary, the charge conductivities $\sigma^B_\ell$ and the heat conductivity $\sigma^B_\epsilon$ are fully determined by the contribution from the LLL, receiving vanishing contributions from the $n_j > 0$ energy levels:
\begin{align}
 \begin{pmatrix}
  \sigma^B_V \\
  \sigma^B_A \\
  \sigma^B_H
 \end{pmatrix} &= \frac{q}{4\pi^2} \sum_{\lambda_j,\varsigma_j}
 \int_0^\infty \frac{dp_j^z}{E_j(e^{\beta \mathcal{E}_j} + 1)}
 \begin{pmatrix} 
  2\lambda_j p_j^z \\  
  \varsigma_j E_j \\
  p^z_j
 \end{pmatrix},\nonumber\\
 \sigma_B^\epsilon &= \frac{q}{4\pi^2} \sum_{\lambda_j,\varsigma_j} 2 \varsigma_j \lambda_j \int_0^\infty \frac{dp_j^z p_j^z}{e^{\beta\mathcal{E}_j} + 1},
\end{align}
where it is understood that $n_j = 0$ in the above relations.

It is worth stressing that the property that only the single lowest Landau level determines the value of a current, while the degenerate higher level contributions mutually cancel each other, is the characteristic feature of the anomalous currents. At the level of the charged currents, it is an expected feature for the chiral magnetic conductivity $\sigma^B_V$, responsible for the Chiral Magnetic Effect~\cite{Fukushima:2008xe}, and the axial magnetic conductivity $\sigma^B_A$, which gives the Chiral Separation Effect~\cite{Son:2004tq}. These conductivities are determined by the axial anomaly~\eq{eq_axial_anomaly}. What is less expected is that a similar statement is also true for the helical conductivity $\sigma^B_H$, which, apparently, is not connected to the axial anomaly but may be related to another form of an anomaly associated with the helical degrees of freedom~\cite{Ambrus:2019ayb,Ambrus:2019khr}.

In the limit of massless fermions, $E_j = |p_j^z|$ on the LLL and the $p_j^z$ integral can be performed analytically, when 
\begin{align}
 \int_0^\infty \frac{dp_j^z}{e^{\beta \mathcal{E}_j} + 1} &= \frac{1}{\beta} \ln\left(1 + e^{\beta \mathbf{q}_j \cdot \bs{\mu}}\right),\nonumber\\
 \int_0^\infty \frac{dp_j^z\, p_j^z}{e^{\beta \mathcal{E}_j} + 1} &= -\frac{1}{\beta^2} {\rm Li}_2(-e^{\beta \bs{q}_j \cdot \bs{\mu}}),
\end{align}
with $\mathbf{q}_j \cdot \bs{\mu} = \varsigma_j \mu_V + 2\lambda_j \varsigma_j \mu_H + 2\lambda_j \mu_A$ and ${\rm Li}_n(z) = \sum_{k = 1}^\infty z^k / k^n$ being the polylogarithm function \cite{DLMF}. Thus, in the high-temperature limit we obtain the following conductivities for massless fermions:
\begin{subequations}\label{eq_sigmas}
\begin{align}
 \sigma^B_V &= \frac{q \mu_A}{2\pi^2} + 
 \frac{q \beta \mu_V \mu_H}{4\pi^2} [1 + O(\beta^2)],
 \label{eq_sigma_V}\\
 \sigma^B_A &= \frac{q \mu_V}{2\pi^2} + 
 \frac{q\beta \mu_A \mu_H}{4\pi^2} [1 + O(\beta^2)], 
  \label{eq_sigma_A}\\
 \sigma^B_H &= 
 \frac{q }{\pi^2 \beta} \ln 2 
 + \frac{q \beta \bs{\mu}^2}{8\pi^2} + O(\beta^3),
 \label{eq_sigma_H}
\end{align}
\end{subequations}
where the higher-order terms appearing in $\sigma^B_V$ and $\sigma^B_A$ vanish when $\mu_V \mu_H = 0$ and $\mu_A \mu_H = 0$, respectively. It is remarkable that the helicity current appears even in a neutral plasma. 

Let us now discuss the conductivities~\eq{eq_sigmas} in more detail. The meaning of these formulas is straightforward: in the background of a magnetic field ${\bs B}$, the fermionic system produces electric ($\ell = V$), axial ($\ell = A$), and helical ($\ell = H$) currents [cf. Eq.~\eq{eq_j_B}],
\begin{align}
{\bs j}_\ell = \sigma^B_\ell {\bs B}\,, \qquad\ \ell = V,A,H\,,
\label{eq_j_B2}
\end{align}
with the appropriate conductivities~\eq{eq_sigmas}.

The first term in Eq.~\eq{eq_sigma_V} gives us the well-known expression for the chiral magnetic conductivity for the chiral separation effect~\eq{eq_CSE} in the absence of the helical chemical potential, $\mu_H = 0$. In particular, the coefficient $1/(2\pi^2)$ in this equation is determined by the axial anomaly which dictates, in turn, the non-conservation of the axial charge in the electromagnetic background:
\begin{align}
\partial_\mu j_A^\mu = \frac{q^2}{16 \pi^2} \epsilon^{\mu\nu\alpha\beta} F_{\mu\nu} F_{\alpha\beta}\,,
 \label{eq_axial_anomaly}
\end{align}
where $\epsilon^{\mu\nu\alpha\beta}$ is the Levi-Civita tensor with $\epsilon^{0123} = +1$. The same statement is true for the first term in the axial magnetic conductivity~\eq{eq_sigma_A} which determines the chiral separation effect~\eq{eq_CSE}. The next-to-leading terms in both these expressions, Eqs.~\eq{eq_sigma_V} and \eq{eq_sigma_A}, give the new high-temperature contributions from the helical imbalance to the appropriate conductivities which are suppressed by the first power of temperature, $O(1/T)$. These conductivities vanish in the electrically neutral and chirally neutral plasmas, at $\mu_V = \mu_A = 0$.

Surprisingly, the helical magnetic conductivity~\eq{eq_sigma_H} differs substantially from its known CME~\eq{eq_sigma_V} and CSE~\eq{eq_sigma_A} counterparts. First of all, the coefficient of proportionality does not show an obvious association with the axial anomaly~\eq{eq_axial_anomaly} due to the presence of the $\ln 2$ factor (see, however, the discussion on suspected role of a new helical anomaly in Refs.~\cite{Ambrus:2019ayb,Ambrus:2019khr}). Secondly, the generation of the helical current is possible in the absence of any (electric, axial or helical) imbalance with all chemical potential set to zero. This property makes it somewhat similar to the mixed axial-gravitational contribution to the heat conductivity which also operates in neutral matter being proportional to the (second power of) temperature~\cite{Landsteiner:2011cp}. The helical conductivity is given by the first term in Eq.~\eq{eq_sigma_H} proportional to temperature $T = 1/\beta$ and we recover our principal result~\eq{eq_HSE}. Finally, similarly to the CME and the CSE effects, the Helical Separation Effect~\eq{eq_HSE} does not change sign under time reversal, $t \to - t$, which is a characteristic feature of anomalous dissipationless transport.

The magnetic heat conductivity evaluates to
\begin{multline}
 \sigma^B_\epsilon = \frac{q \mu_H}{\pi^2 \beta} \ln 2 + 
 \frac{q \mu_V \mu_A}{2\pi^2} \\
 + \frac{q \beta \mu_H}{24 \pi^2} \left[3 \bs{\mu}^2 - 2\mu_H^2 + O(\beta^2)\right],
\label{eq_heat_conductivity}
\end{multline}
where the higher-order terms vanish when $\mu_H = 0$. In deriving the above expression, we employed the following expansion of the polylogarithm:
\begin{equation}
 {\rm Li}_2(-e^{x})= -\frac{\pi^2}{12} - x \ln 2 - \frac{x^2}{4} - \frac{x^3}{24} + \frac{x^5}{960} + O(x^7).
\end{equation}

As mentioned already, the magnetic conductivities related to anomalous transport receive contributions only from the LLL. 
The other quantities appearing in Eqs.~\eqref{eq:thermo_macro}--\eqref{eq:thermo_piB} receive contributions also from the HLL.
However, in the limit of a strong magnetic field, $|qB|\gg T^2$, the HLL contributions are exponentially suppressed. Moving as before to the case of massless fermions, we find
\begin{align}
 Q_\ell &= \frac{|qB| \mu_\ell}{2\pi^2} + 
 \frac{|qB| \beta}{4\pi^2} \frac{\partial \mu_\times^3}{\partial \mu_\ell} \left[1 + O(\beta^2)\right],\nonumber\\
 \epsilon &= 3(P + \Pi) = \frac{3}{2} \pi_B = \frac{|qB|}{12\beta^2} \nonumber\\
 & + \frac{|qB| \bs{\mu}^2}{4\pi^2} + \frac{|qB| \beta}{4\pi^2} \mu_\times^3 \left[1 + O(\beta^2)\right],
 \label{eq:thermo_macro_LLL_m0}
\end{align}
where we denoted $\bs{\mu}^2 = \mu_V^2 + \mu_A^2 + \mu_H^2$ and $\mu_\times^3 = \mu_V \mu_A \mu_H$.

\subsection{Thermodynamic analysis}
\label{sec:anomalous_transport:thermo}

Let us consider the grand canonical potential of the system, 
\begin{equation}
 \Phi = -\frac{|qB| V}{4\pi^2 \beta} \sum_{n_j = 0}^\infty \sum_{\lambda_j, \varsigma_j} \int_{0}^\infty dp_j^z \, g_j \ln(1 + e^{-\beta \mathcal{E}_j}),
\end{equation}
where $V$ is the (infinite) system size.
The grand potential $\Phi$ is related to the partition function $\mathcal{Z} = {\rm tr}(\hat{\rho})$ via
\begin{equation}
 \mathcal{Z} = e^{-\beta \Phi}.
\end{equation}

Given the thermodynamic relation 
\begin{equation}
 d\Phi = \beta^{-2} \mathcal{S} d\beta - P dV - \bs{\mathcal{Q}} \cdot d\bs{\mu},
\end{equation}
we can identify the total entropy $\mathcal{S}$, thermodynamic pressure $\mathcal{P}$ and total charge $\bs{\mathcal{Q}}$ as 
\begin{equation}
 \mathcal{S} = \beta^2 \frac{\partial \Phi}{\partial \beta}, \quad 
 \mathcal{P} = -\frac{\partial \Phi}{\partial V}, \quad 
 \mathcal{Q}_\ell = -\frac{\partial \Phi}{\partial \mu_\ell}.
\end{equation}
Since the grand potential is linearly proportional to the system volume, $\Phi = \phi V$ with $\phi$ being a volume-independent quantity, it is clear that 
\begin{equation}
 P = -\phi = \Theta^{zz}.
\end{equation}
It is easy to check that $Q_{V/H} \equiv \mathcal{Q}_{V/H} / V = \partial \Theta^{zz} / \partial \mu_{V/H}$. A similar relation holds for $Q_A$ only in the case of massless fermions ($M = 0$ and $\mathfrakE^+_j = \mathfrakE^-_j = 1$), confirming that the chiral charge is conserved only for massless fermions. 

The total energy $\mathcal{E} = \langle \widehat{H} \rangle = \mathcal{Z}^{-1} {\rm tr}(\hat{\rho} \widehat{H})$ can be obtained from $\mathcal{Z}$ via 
\begin{equation}
 \mathcal{E} = \bs{\mu} \cdot \bs{\mathcal{Q}} - \frac{\partial \ln \mathcal{Z}}{\partial \beta}.
\end{equation}
Since $\ln \mathcal{Z} = -\beta \Phi$, it is easy to uncover the Euler relation 
\begin{equation}
 s = \beta(\epsilon + P - \bs{\mu} \cdot \bs{Q}),
\end{equation}
where $s = \mathcal{S} / V$, $\epsilon = \mathcal{E} / V$ and $\bs{Q} = \bs{\mathcal{Q}} / V$. 

The above discussion shows that the entropy density can be obtained as
\begin{equation}
 s = \beta\left(\Theta^{tt} + \Theta^{zz} - \bs{\mu} \cdot \frac{\partial \Theta^{zz}}{\partial \bs{\mu}}\right) = \frac{\partial \Theta^{zz}}{\partial T}.
\end{equation}
Having identified $P = \Theta^{zz}$, we are in a position to evaluate the dynamical pressure $\Pi$ using Eq.~\eqref{eq:hydro_eps_P_plus_Pi}:
\begin{align}
 \Pi &= \frac{1}{3}(\Theta^{xx} + \Theta^{yy} - 2\Theta^{zz}) = -\pi_B \nonumber\\
 &= \frac{|qB|}{6\pi^2} \sum_{n_j = 0}^\infty \sum_{\lambda_j, \varsigma_j}
 \int_0^\infty \frac{dp_j^z\,g_j (n_j |qB| - p_{z;j}^2)}{E_j(e^{\beta \mathcal{E}_j} + 1)}. 
 \label{eq_Pi}
\end{align}

On the LLL and for massless fermions, the relation $\Pi = -\pi_B$ can be used in conjunction with Eq.~\eqref{eq:thermo_macro_LLL_m0} to obtain:
\begin{equation}
 \epsilon = P, \quad \pi_B = -\Pi = \frac{2}{3} P.
\end{equation}

\section{Search for helical excitations in neutral plasma}
\label{sec_helical_heat}

Let us consider a globally-neutral plasma, characterized by vanishing chemical potentials ${\bar \mu}_\ell = 0$ ($\ell = V,A,H$) and temperature ${\bar T}$, where the bar over the symbol indicates that we consider a mean global coordinate-independent quantity. We work in the hydrodynamic approximation, which assumes small departures from local thermodynamic equilibrium.
We also assume the limit of a strong magnetic field, $|q B| \gg T^2$.

In a strong magnetic field, the excess in vector or axial densities generates a coherent propagation of these charges along the direction of the magnetic field in the form of a linear hydrodynamic excitation, known as the Chiral Magnetic Wave~\cite{Kharzeev:2010gd}. In this wave, an excess in vector (axial) charge generates an axial (vector) current along the magnetic field, which leads to a build-up of an axial (vector) charge and a cyclic repetition of the process.

We look for a similar excitation in terms of helical degrees of freedom. To this end we notice that the helical current ${\bs j}_H = \sigma^B_H {\bs B}$ and the heat current ${\bs j}_\epsilon = \sigma^B_\epsilon {\bs B}$, can be expressed, respectively, via Eq.~\eq{eq_sigma_H} for the helical conductivity~$\sigma^B_H$ and Eq.~\eq{eq_heat_conductivity} for the heat conductivity~$\sigma^B_\epsilon$. These relations show that helical fluctuations are coupled to the fluctuations in temperature similarly to the coupling of vector and axial charges (currents) in the Chiral Magnetic Wave. Therefore, one could expect that the helical degrees of freedom may propagate in terms of a hypothetical ``helical heat wave,'' which would corresponds to a coherent hydrodynamic excitation that combines energy density and helical fluctuations.

To assess the existence of the helical heat wave as a linear hydrodynamic excitation, we consider the equations of motion derived from the charge (for all $\ell = V,A,H$), energy, and momentum conservation, namely:
\begin{subequations}
\begin{align}
 \dot{Q}_\ell + Q_\ell \theta + \partial_\mu j^\mu_\ell &= 0, \label{eq:hydro_charge}\\
 \dot{\epsilon} + (\epsilon + P + \Pi) \theta - \pi^{\mu\nu} \sigma_{\mu\nu} + \partial_\mu j_\epsilon^\mu - j_\epsilon^\nu \dot{u}_\nu &= 0,\label{eq:hydro_eps}\\
 (\epsilon + P + \Pi) \dot{u}^\mu - \nabla^\mu(P + \Pi) + \Delta^\mu_\lambda \partial_\nu \pi^{\nu \lambda} & \nonumber\\
 + j_\epsilon^\lambda \nabla_\lambda u^\mu + Dj_\epsilon^{\langle \mu \rangle} + j_\epsilon^\mu \theta &= 0, \label{eq:hydro_mom}
\end{align}
\end{subequations}
where $D$ and the overhead dot denote the comoving derivative, $Da \equiv \dot{a} \equiv u^\mu \partial_\mu a$,
$\sigma_{\mu\nu} = \Delta^{\alpha\beta}_{\mu\nu} \partial_\alpha u_\beta$ is the shear tensor with $\Delta^{\alpha\beta}_{\mu\nu}$ introduced in Eq.~\eq{eq:Delta4} and $\theta = \partial_\mu u^\mu$ is the expansion scalar. In the following, we focus on the dissipationless transport phenomena supported by the above set of equations and ignore dissipative corrections to the non-ideal terms $\Pi$, $\pi^{\mu\nu}$, $j_\epsilon^\mu$, and $j_\ell^\mu$. Also, we consider that the electric field is negligible. In reality, one should allow for fluctuations in $B^\mu$ and in the electric field $E^\mu$, coming from the Maxwell equations. The backreaction from fluctuating electromagnetic field usually leads to a damping effect~\cite{Kharzeev:2010gd} which we neglect below.

We now consider infinitesimal fluctuations in a quiescent, neutral fluid of massless particles, at background temperature $\overline{T}$, in a strong magnetic field $|qB| \gg \overline{T}^2$. We consider the generic split of a hydrodynamic quantity $a$ into its global average $\bar{a}$ and fluctuation $\delta a$ as follows,
\begin{equation}
 a = \bar{a} +\delta a, \quad 
\delta a = \int d^3k \widetilde{\delta a}_\bk(\omega) e^{-i\omega t + i \bk \cdot \bx},
\label{eq:hydro_Fourier}
\end{equation}
where $\widetilde{\delta a}_\bk(\omega)$ represents the amplitude of its Fourier mode of wave vector $\mathbf{k}$. For brevity, we henceforth suppress the explicit $\bk$ and $\omega$ dependencies of fluctuations.

We take as independent parameters the fluid temperature $T$, chemical potentials $\mu_\ell$, and macroscopic velocity $u^\mu$, with $\bar{\mu}_\ell = 0$ and $\bar{u}^\mu = \delta^\mu_0$.
The velocity fluctuations lead to a modification $u^\mu = \bar{u}^\mu + \delta u^\mu$ that affects also the covariant magnetic field $B^\mu = \overline{B}^\mu + \delta B^\mu$, such that:
\begin{equation}
 \overline{B}^\mu = B \delta^\mu_z, \quad 
 \delta B^\mu = B \delta u^z \delta^\mu_z.
\end{equation}
Also, the comoving derivative $\dot{a}$ becomes 
\begin{equation}
 \dot{a} \simeq \partial_t \delta a \rightarrow 
 -i \omega \widetilde{\delta a},
\end{equation}
where the arrow indicates that the Fourier transform was taken as in Eq.~\eqref{eq:hydro_Fourier}. More generally, the derivative leads simply to the replacement
\begin{equation}
 \partial_\mu a^\nu \rightarrow -i k_\mu \widetilde{\delta a^\nu},
\end{equation}
and in particular, $\theta \rightarrow -i k_\mu \widetilde{\delta u^\mu} = i \mathbf{k} \cdot \widetilde{\delta \mathbf{u}}$. 

In the case of the charge conservation equation, Eq.~\eqref{eq:hydro_charge}, we have
\begin{align}
 \overline{Q}_\ell &= 0, & \delta Q_\ell &= \frac{|qB|}{2\pi^2} \delta \mu_\ell, \nonumber\\
 \bar{\sigma}^B_{V/A} &= 0, & 
 \delta \sigma^B_{V/A} &= \frac{q}{2\pi^2} \delta \mu_{A/V},\nonumber\\
 \bar{\sigma}^B_H &= \frac{q \overline{T} \ln 2}{\pi^2}, & 
 \delta \sigma^B_H &= \frac{q \ln 2}{\pi^2} \delta T.
\end{align}
Taking into account that $\bar{\theta} = 0$, the second term $Q_\ell \theta$ in Eq.~\eqref{eq:hydro_charge} becomes negligible. The third term evaluates to
\begin{align}    
 \partial_\mu j^\mu_\ell & = B \partial_z \delta \sigma^B_\ell + B \bar{\sigma}^B_\ell \partial_t \delta u^z \nonumber \\
  & \rightarrow 
 ik^z B \widetilde{\delta \sigma}^B_\ell -i \omega B \bar{\sigma}^B_\ell \widetilde{\delta u^z}.
\end{align}
Since $\bar{\sigma}^B_{V/A} = 0$, the equations giving the conservation of $Q_V$ and $Q_A$ reduce to the closed set
\begin{equation}
\begin{pmatrix}
 \omega & -k^z \sigma \\
 -k^z \sigma & \omega 
\end{pmatrix}
\begin{pmatrix}
\widetilde{\delta \mu}{}_V \\ \widetilde{\delta \mu}{}_A 
\end{pmatrix} = 0.
\label{eq:hydro_VA}
\end{equation}
The above equations are solved when either $\omega = \pm k^z$ or when both $\widetilde{\delta \mu}{}_V$ and $\widetilde{\delta \mu}{}_A$ cancel. Hence, we have uncovered the chiral magnetic wave which comprises coherent oscillations of vector and axial charges propagating, in strong magnetic field background, with the speed of light~\cite{Kharzeev:2010gd}. 

Since $\bar{\sigma}^B_H \neq 0$, the conservation equation for the helicity charge $Q_H$ leads to a coupling between the helicity chemical potential $\widetilde{\delta \mu}{}_H$ and the fluctuations in the hydrodynamic sector,
\begin{equation}
 \omega \widetilde{\delta u}{}^z - k^z \frac{\widetilde{\delta T}}{\overline{T}} + 
 \frac{\sigma \omega}{2\ln 2} \frac{\widetilde{\delta \mu}{}_H}{\overline{T}} = 0.
 \label{eq:hydro_H_Fourier}
\end{equation}

Moving now to the $T^{\mu\nu}$ sector, Eq.~\eqref{eq:thermo_macro_LLL_m0} shows that $\epsilon = P$ and $\Pi = -\pi_B = -\frac{2}{3} P$. Considering that these equalities hold also in the perturbed system, we have 
\begin{align}
 \bar{\epsilon} &= \frac{|qB| \overline{T}^2}{12}, &
 \delta \epsilon &= \frac{|qB| \overline{T}}{6} \delta T,\nonumber\\
 \bar{\sigma}^B_\epsilon &= 0, & 
 \delta \sigma^B_\epsilon &= \frac{q \overline{T} \ln 2}{\pi^2} \delta \mu_H.
\end{align}

In the case of the shear-stress tensor $\pi^{\mu\nu}$, we ignore dissipative corrections and employ the form in Eq.~\eqref{eq:hydro_pimunu}. Writing $\pi^{\mu\nu} = \bar{\pi}^{\mu\nu} + \delta \pi^{\mu\nu}$, we have $\bar{\pi}^{\mu\nu} = \frac{1}{2} \bar{\pi}_B \times {\rm diag}(0, -1,-1,2)$ with $\bar{\pi}_B = \frac{2}{3} \overline{P}$ and 
\begin{multline}
 \delta \pi^{\mu\nu} = \frac{1}{2} \delta \pi_B {\rm diag}(0,-1,-1,2) \\ 
 + \frac{\bar{\pi}_B}{2} 
 \begin{pmatrix}
  0 & - \delta u^x & -\delta u^y & 2\delta u^z \\
  -\delta u^x & 0 & 0 & 0\\
  -\delta u^y & 0 & 0 & 0 \\
  2\delta u^z & 0 & 0 & 0
 \end{pmatrix}.
\end{multline}
The shear tensor $\sigma^{\mu\nu}$ has vanishing average, $\bar{\sigma}^{\mu\nu} = 0$, while $\delta \sigma^{\mu\nu} = \nabla_{(\mu} \delta u_{\nu)} - \frac{1}{3} \Delta_{\mu\nu} \theta$ has vanishing components when either $\mu = t$ or $\nu = t$, i.e. $\delta \sigma^{00} =\delta \sigma^{0i} = \delta \sigma^{i0} = 0$. On the spatial part, we have $\delta \sigma_{ij} = \partial_{(i} \delta u_{j)} - \frac{1}{3} g_{ij} \theta$, or in Fourier space,
\begin{align}
 \widetilde{\delta \sigma}^{ij} = -\frac{i}{2} (k^i \widetilde{\delta u}^j + k^j \widetilde{\delta u}^i) - \frac{i}{3} g^{ij} \mathbf{k} \cdot \widetilde{\delta \mathbf{u}}.
\end{align}

In the case of the energy conservation equation, we employ $\pi^{\mu\nu} \sigma_{\mu\nu} = -\frac{1}{2} \bar{\pi}_B(\delta \sigma^{xx} + \delta \sigma^{yy} - 2\delta \sigma^{zz}) \rightarrow \frac{i}{3} \overline{P} (\mathbf{k} \cdot \widetilde{\delta \mathbf{u}} - 3 k^z \widetilde{\delta u^z})$, $j_\epsilon^\nu \dot{u}_\nu \simeq 0$, and 
\begin{equation}
 \partial_\mu j_\epsilon^\mu = B \partial_z \delta \sigma^B_\epsilon \rightarrow \frac{ik \overline{P}}{\overline{T}} \frac{12 \ln 2}{\pi^2} \widetilde{\delta \mu}_H
\end{equation}
to arrive at
\begin{equation}
 2\omega \frac{\widetilde{\delta T}}{\overline{T}} - \bk \cdot \widetilde{\delta \mathbf{u}} - k^z \widetilde{\delta u}{}^z - \frac{12 \sigma \ln 2}{\pi^2} k^z \frac{\widetilde{\delta \mu}{}_H}{\overline{T}} = 0.
 \label{eq:hydro_eps_Fourier}
\end{equation}

Now we move on to the momentum conservation equation, i.e. Eq.~\eqref{eq:hydro_mom}. 
The terms $j_\epsilon^\lambda \nabla_\lambda u^\mu$ and $j_\epsilon^\mu \theta$ are of higher order and can thus be ignored. Furthermore, the projector $\Delta^\mu_\nu$ together with $\dot u^0 \simeq 0$ ensures that the $\mu = 0$ relation reduces to a tautological $0 = 0$.
For $\mu = i \in \{x, y\}$ and $\mu = z$, we have
\begin{align}
 \Delta^i_\lambda \partial_\nu \pi^{\nu \lambda} &= \frac{1}{2} \partial^i \delta \pi_B + \frac{\bar{\pi}_B}{2} \partial_t \delta u^i \rightarrow -\frac{i\overline{P}}{3} \left(2 k^i \frac{\widetilde{\delta T}}{\overline{T}} - \omega \widetilde{\delta u^i}\right), \nonumber\\
 \Delta^z_\lambda \partial_\nu \pi^{\nu \lambda} &= \partial_z \delta \pi_B + \bar{\pi}_B \partial_t \delta u^z 
 \rightarrow \frac{2i \overline{P}}{3} \left(2 k^z \frac{\widetilde{\delta T}}{\overline{T}} - \omega \widetilde{\delta u^z}\right).
\end{align}
We thus arrive at:
\begin{align}
 \omega \widetilde{\delta u}{}^i &= 0, & 
 \omega \widetilde{\delta u}{}^z - k^z \frac{\widetilde{\delta T}}{\overline{T}} + \omega \frac{6 \sigma \ln 2}{\pi^2} \widetilde{\delta \mu}{}_H &= 0. 
\label{eq:hydro_mom_Fourier}
\end{align}
The first relation above shows that the plasma does not support oscillations in the plane orthogonal to the magnetic field. The second relation in Eq.~\eqref{eq:hydro_mom_Fourier} can be used in conjunction with Eqs.~\eqref{eq:hydro_H_Fourier} and \eqref{eq:hydro_eps_Fourier} to derive 
\begin{align}
 \begin{pmatrix}
  \omega & -k^z & {\displaystyle \frac{\sigma\omega}{2\ln 2}} \smallskip\\
  -k^z & \omega & {\displaystyle -\frac{6 \sigma \ln 2}{\pi^2} k^z} \smallskip\\
\omega & -k^z & {\displaystyle \frac{6 \sigma \ln 2}{\pi^2} \omega}
\end{pmatrix}
\begin{pmatrix}
 \widetilde{\delta u}{}^z \smallskip\\
 {\displaystyle \frac{\widetilde{\delta T}}{\overline{T}}} \smallskip\\
 {\displaystyle \frac{\widetilde{\delta \mu}{}_H}{\overline{T}}}
\end{pmatrix} = 0.
\end{align}
In order to have non-trivial solutions of the above equation, the determinant of the matrix on the left-hand side must vanish. This leads to the equation
\begin{equation}
 \sigma \omega\left(\frac{1}{ 2\ln 2} - \frac{6\ln 2}{\pi^2} \right) (\omega^2 - k_z^2) = 0,
\end{equation}
where the solution $\omega = 0$ corresponds to the mode that supports helicity chemical potential fluctuations. The gapless modes $\omega = \pm k^z$ have identically $\widetilde{\delta \mu}_H = 0$ and correspond to the standard sound modes present in the ideal fluid, which in the present case propagate at the speed of light.

\section{Conclusion}
\label{sec:conc}

In our paper, we have found the Helical Separation Effect (HSE) which implies that charged massless fermions develop a dissipationless flow of helicity along the background magnetic field~\eq{eq_HSE}. This dissipationless effect is similar to the Chiral Separation Effect (CSE)~\cite{Son:2004tq}, which generates the flow of the chiral charge along magnetic field lines in dense fermionic matter. Contrary to the CSE, the HSE produces the helical current in a neutral plasma in which all chemical potentials vanish. Moreover, the expression for the HSE current~\eq{eq_HSE} contains a ``$\ln 2$'' factor thus making it difficult to associate this effect to a known anomaly (see, however, the discussions in Ref.~\cite{Ambrus:2019ayb,Ambrus:2019khr} on possible existence of helical anomalies). The associated helical conductivity is determined by temperature which makes it distantly similar to the axial-gravitational effects~\cite{Landsteiner:2011cp}. The presence of matter plays an auxiliary role enhancing the helical current~\eq{eq_HSE_2}, \eq{eq_kappa_beta} generated by the HSE. The same remarks are also valid for the Helical Magnetic Heat Effect~\eq{eq_heat_flux} which generates, in the presence of non-vanishing helical charge density, the fermionic heat flux along the magnetic field.

Since the helical current generated by the HSE is linearly proportional to the background temperature and magnetic field, the helical current~\eq{eq_HSE} and the heat current~\eq{eq_heat_flux} become coupled together via the HSE in a striking similarly with the coupling -- mediated by the CME and CSE -- of the vector and chirality currents that result in the appearance of the chiral magnetic wave~\cite{Kharzeev:2010gd}. However, a detailed analysis shows that the fluctuations of the helical charge in a neutral plasma do not lead to the appearance of a new hydrodynamic mode. The helical degree of freedom may, however, lead to the appearance of additional hydrodynamic waves in dense plasma, which are left beyond the scope of this paper.

The helicity plays a complementary role to chirality, as the total helicity of a particle-antiparticle ensemble is given by the sum of axial charges carried by particles minus the sum of the axial charges possessed by antiparticles. As the helicity is different from chirality, its role in phenomenological applications in Quark-Gluon plasma may lead  to new observational effects which remain to be explored.

\acknowledgments
The authors are grateful to Yago Ferreiros for collaborating on an earlier related project. VEA gratefully acknowledges the support through a grant of the Ministry of Research, Innovation and Digitization, CNCS - UEFISCDI, project number PN-III-P1-1.1-TE-2021-1707, within PNCDI III. The authors gratefully acknowledge support by the European Union - NextGenerationEU through the grant No. 760079/23.05.2023, funded by the Romanian ministry of research, innovation and digitalization through Romania’s National Recovery and Resilience Plan, call no. PNRR-III-C9-2022-I8.

\bibliography{HSE.bib}

\end{document}